\documentclass[12pt]{article}
\usepackage{amsmath}
\usepackage{amssymb}
\usepackage{slashed}
\usepackage[square,numbers,sectionbib,compress]{natbib}
\usepackage{graphicx}
\usepackage{breqn}
\usepackage{bbold}
\usepackage{color}
\allowdisplaybreaks

\newcommand{\eps}{\varepsilon}

\newcommand{\D}{\mathcal{D}}
\newcommand{\Op}{\mathcal{O}}
\newcommand{\G}{\mathcal{G}}
\newcommand{\CA}{$C_A$}
\newcommand{\CF}{$C_F$}
\newcommand{\nf}{$n_f$}

\usepackage{pslatex}
\usepackage[latin1]{inputenc}
\usepackage[T1]{fontenc}

\makeatletter
\DeclareRobustCommand{\cev}[1]{%
  {\mathpalette\do@cev{#1}}%
}
\newcommand{\do@cev}[2]{%
  \vbox{\offinterlineskip
    \sbox\z@{$\m@th#1 x$}%
    \ialign{##\cr
      \hidewidth\reflectbox{$\m@th#1\vec{}\mkern4mu$}\hidewidth\cr
      \noalign{\kern-\ht\z@}
      $\m@th#1#2$\cr
    }%
  }%
}
\makeatother

\begin{document}

\numberwithin{equation}{section}

\begin{titlepage}
\noindent
\hfill August 2023\\
\vspace{0.6cm}
\begin{center}
{\LARGE \bf 
    Basis transformation properties of anomalous dimensions for hard exclusive processes}\\ 
\vspace{1.4cm}

\large
S.~Van Thurenhout$^{\, a}$\\
\vspace{1.4cm}
\normalsize
{\it $^{\, a}$HUN-REN Wigner Research Centre for Physics, Konkoly-Thege Mikl\'os u. 29-33, 1121 Budapest, Hungary}\\
\vspace{1.4cm}

{\large \bf Abstract}
\vspace{-0.2cm}
\end{center}
When considering the renormalization of composite operators for the description of hard exclusive scattering processes, two types of operator basis called the \textit{derivative basis} and the \textit{Gegenbauer basis} are often used in the literature. In this work we set up the explicit similarity transformations between these two bases, both for quark and gluon operators. This way, one can use the properties of both bases to their advantage in the computation of the operator anomalous dimensions, which describe the scale dependence of non-perturbative non-forward parton distributions. We provide several applications of our framework. As an application of the gluon transformation formula, we compute the one-loop non-forward purely gluonic anomalous dimension matrix. For the rest of the applications we focus on the transformation formula of the quark operator. We derive the Gegenbauer anomalous dimensions, in the limit of a large number of quark flavors $n_f$, to all orders in the strong coupling $\alpha_s$, extending the computation previously performed in the derivative basis. Next a numeric calculation of the two-loop anomalous dimensions in the derivative basis beyond the leading-color limit is presented. Finally, we discuss a novel way of validating existing computations of the conformal anomaly based on the leading-color anomalous dimensions in the derivative basis. 
\vspace*{0.3cm}
\end{titlepage}

\section{Introduction}
The study of exclusive processes, like for example deeply virtual Compton scattering, provides useful information relating to the inner structure of the proton. Such information is encoded in non-perturbative objects like the generalized parton distributions (GPDs)~\cite{Muller:1994ses,Ji:1996ek,Ji:1996nm,Radyushkin:1996nd,Radyushkin:1996ru,Diehl:2003ny} and is vital in the quest of understanding the emergence of hadronic properties from the partonic ones, e.g. the emergence of the proton spin from spins and angular momenta of its constituent partons~\cite{Aidala:2012mv,Leader:2013jra,Deur:2018roz,Ji:2020ena}. Experimental studies in the past were performed e.g. by the {\sc HERA} collaboration~\cite{Abramowicz:2015mha,Accardi:2016ndt}, and the understanding of GPDs will be a major goal of a future Electron Ion Collider (EIC)~\cite{Boer:2011fh,AbdulKhalek:2021gbh}.\newline

On the theoretical side, insight into the non-perturbative parton distributions is gained from studying hadronic matrix elements of composite operators. For the exclusive processes of interest, these matrix elements are non-forward, i.e. there is a non-zero momentum running through the operator vertex. As such, there will be mixing with total-derivative operators. This is particularly important in the renormalization procedure of the operators, which is necessary to remove the ultraviolet divergences. When renormalizing the partonic matrix elements of the operators, which can be done perturbatively in the strong coupling $\alpha_s$, one gains access to the elements of the operator anomalous dimension matrix (ADM). The latter are important for phenomenology as they determine the energy scale dependence of the non-perturbative parton distributions. In particular, the diagonal elements of the ADM correspond to the forward anomalous dimensions and determine the scale dependence of forward parton distributions like the standard PDFs. These have been studied in depth, with partial information available to order $\alpha_s^5$ \cite{Gross:1973ju,Floratos:1977au,Moch:2004pa,Blumlein:2021enk,Gracey:1994nn,Davies:2016jie,Velizhanin:2011es,Velizhanin:2014fua,Ruijl:2016pkm,Moch:2017uml,Herzog:2018kwj,Blumlein:2023aso}. Likewise, the off-diagonal elements of the ADM, which are only relevant for exclusive scattering processes, correspond to the non-forward anomalous dimensions and determine the scale dependence of non-forward parton distributions like the GPDs. Because of their importance, a lot of effort has gone into the computation of the ADM elements. Using conformal symmetry techniques, the evolution kernel was computed to the three-loop level~\cite{Braun:2003rp,Braun:2017cih}. The reconstruction of the anomalous dimensions in this approach relies on consistency relations coming from the conformal algebra and requires, beyond the one-loop level, the computation of the so-called conformal anomaly. The latter is known to two-loop accuracy, although a closed-form expression for the two-loop quantity is currently not known~\cite{Braun:2017cih,Mueller:1991gd,Braun:2016qlg}. As generically the $L$-loop anomalous dimensions depend only on the ($L-1$)-loop conformal anomaly~\cite{Mueller:1991gd}, they could be calculated up to three loops using this approach.\newline

The anomalous dimensions can also be determined directly by computing and renormalizing the appropriate partonic operator matrix elements. This is relatively straightforward in the case of forward kinematics, for which there is no momentum flow through the operator vertex. In particular, the forward anomalous dimensions (i.e. the diagonal elements of the ADM) are straightforwardly related to the 1/$\eps$-pole of the matrix elements in $D=4-2\eps$ dimensional regularization. The case of non-forward kinematics, for which there is a non-zero momentum flow through the operator vertex, is more complicated because of the mixing with total-derivative operators. Using this direct method, the one-loop off-diagonal elements of the ADM were computed in~\cite{Artru:1989zv,Shifman:1980dk,Baldracchini:1981,Geyer:1982fk,Blumlein:1999sc,Blumlein:2001ca} while low-$N$ fixed moments at orders $\alpha_s^2$ and $\alpha_s^3$ were presented in~\cite{Gracey:2009da,Kniehl:2020nhw}. The direct method was extended in~\cite{Moch:2021cdq,VanThurenhout:2022nmx} to produce general expressions for the off-diagonal elements of the ADM using a consistency relation between the anomalous dimensions following from the renormalization structure of the operators in the chiral limit. This method was used to obtain the anomalous dimensions in the limit of a large number of quark flavors $n_f$ to order $\alpha_s^5$ for Wilson operators and $\alpha_s^4$ for transversity operators. For Wilson operators it was also used to compute the two-loop anomalous dimensions in the leading-color limit.\newline

Finally, all-order results in the large-$n_f$ limit recently became available, both for Wilson and transversity operators~\cite{VanThurenhout:2022hgd}. The method to obtain these hinged on exact conformal symmetry of quantum chromodynamics (QCD) at the Wilson-Fisher fixed point and extends similar computations for the forward anomalous dimensions~\cite{Gracey:1994nn,Gracey:2003mr}.\newline

As already mentioned previously, the determination of the anomalous dimensions in non-forward kinematics is non-trivial because of operator mixing. However, one can make significant simplifications by choosing an appropriate basis for the total-derivative operators. The conformal method described above naturally works in the so-called \textit{Gegenbauer basis}, while the consistency relation between the anomalous dimensions is valid in what is called the \textit{derivative basis}. In~\cite{Moch:2021cdq} we derived an implicit formula to check the consistency of results in both bases. Implicit here means that sums of anomalous dimensions in one basis were related to sums of anomalous dimensions in the other basis. In this work, we extend this to an explicit similarity transformation between the two bases, which allows one to take advantage of the properties of both bases. The rest of the text consists of applications of our newfound transformation.\newline

We will focus our analysis below on the leading-twist operators, which provide the dominant effects in an expansion in inverse powers of the hard scale of the scattering process. Higher-twist corrections are becoming increasingly important however due to the high accuracy of experimental data and planned future experiments at lower energies (e.g. the EIC will have a maximum center-of-mass energy of around 140 GeV). Furthermore, certain physical observables, like for example single spin asymmetries \cite{Anselmino:1994gn,Liang:2000gz,Barone:2001sp,Braun:2009mi}, are not accessible in the leading-twist approximation. For an overview of recent progress in higher-twist physics, we refer the reader to \cite{Braun:2022gzl}.
\newline

The article is organized as follows: In Sec.~\ref{sec:theory} we introduce the leading-twist quark operators of interest and give a technical overview of the operator bases. In Sec. \ref{sec:trans} we present the derivation of the basis transformation formulae, both for quark and gluon operators, and discuss some generic consequences. The rest of the article deals with applications of our newly derived formulae. Sec.~\ref{sec:largeNF} extends the all-order large-$n_f$ anomalous dimensions in the derivative basis of~\cite{VanThurenhout:2022hgd} to those in the Gegenbauer basis. Numeric results in QCD up to sixth order in the strong coupling will be presented for both Wilson and transversity operators up to spin five. The corresponding all-order results are collected in Appendix \ref{sec:appA}. In the next section we summarize the analytic status of the two-loop anomalous dimensions in the derivative basis beyond the leading-color approximation, and present numerical results for the \textit{full} spin-ten two-loop anomalous dimensions. The corresponding values of the $SU(n_c)$ anomalous dimensions are collected in Appendices \ref{ap:L2ADM} and \ref{ap:L2ADMTr}. Sec.~\ref{sec:anomaly} discusses a new approach to cross-check existing computations of the so-called conformal anomaly, which is an important ingredient for the calculation of operator anomalous dimensions in conformal schemes. Finally in Sec.~\ref{sec:gluon}, we present the non-forward fully gluonic ($\gamma^{\:gg}$) anomalous dimensions to order $\alpha_s$. Conclusions and an outlook are given in Sec.~\ref{sec:conclusion}.

\section{Theoretical framework}
\label{sec:theory}
We are interested in the renormalization of the flavor-non-singlet quark operators in QCD, which are generically defined as
\begin{equation}
\label{eq:operators}
    \mathcal{O} = \mathcal{S}\overline{\psi}\lambda^{\alpha}\Gamma D_{\nu_2}\dots D_{\nu_N} \psi.
\end{equation}
Here $D_{\mu}=\partial_{\mu}-ig_sA_{\mu}$ is the QCD covariant derivative and $\lambda^{\alpha}$ the generators of the $SU(n_f)$ flavor group. Depending on the physical process different Dirac structures have to be considered in the operators of Eq.(\ref{eq:operators}). We generically denote this by $\Gamma$. The following two operator types will be discussed in this work:
\begin{itemize}
    \item $\Gamma=\gamma_{\nu_1}$ corresponding to Wilson operators and
    \item $\Gamma=\sigma_{\mu\nu_1}\equiv \frac{1}{2}[\gamma_{\mu},\gamma_{\nu_1}]$ for transversity operators.
\end{itemize} 
Furthermore we will only consider the leading-twist operators, meaning we symmetrize the Lorentz indices $\nu_1\dots\nu_N$ and subtract the traces, which is denoted by $\mathcal{S}$. 
These operators are important phenomenologically, as their hadronic matrix elements define the non-perturbative parton distributions appearing in the QCD factorized hard scattering cross sections. As such, the scale dependence of the distributions is set by that of the operators, the latter being determined by the operator anomalous dimension,
\begin{equation}
\label{eq:defGam}
    \frac{\text{d}[\mathcal{O}]}{\text{d}\ln\mu^2} = -\gamma[\mathcal{O}].
\end{equation}
The brackets $[\:]$ signify that the operators are renormalized. Hence the scale dependence of the distributions, contrary to the distributions themselves, can be calculated perturbatively in the strong coupling $a_s=\alpha_s/(4\pi)$,
\begin{equation}
    \gamma \equiv a_s \gamma^{(0)} +a_s^2 \gamma^{(1)} + \dots.
\end{equation}
When studying hard exclusive scattering processes, mixing of the operators in Eq.(\ref{eq:operators}) with total-derivative ones has to be taken into account. As such, Eq.(\ref{eq:defGam}) becomes a matrix equation and the corresponding anomalous dimensions determine the scale dependence of the corresponding non-forward parton distributions (like e.g. the GPDs) through the ERBL equation~\cite{Efremov:1978rn,Efremov:1979qk,Lepage:1979zb,Lepage:1980fj}. The same anomalous dimensions also appear in the Brodsky-Lepage equation, which characterizes the scale dependence of hadronic wavefunctions \cite{Lepage:1979zb,Lepage:1980fj}
\begin{equation}
\label{eq:ERBL1}
    \frac{\text{d}\phi(x,\mu^2)}{\text{d}\ln\mu^2} = \int_{0}^{1}\text{d}y\:V(x,y)\phi(y,\mu^2)
\end{equation}
with $\phi$ a generic hadronic wavefunction and $\mu$ the renormalization scale. The kernel $V(x,y)$ is related to the elements of the anomalous dimension matrix through~\cite{Dittes:1988xz}
\begin{equation}
\label{eq:ERBL2}
    \sum_{k=0}^N\gamma_{N,k}\:y^k=-\int_{0}^{1}\text{d}x\:x^N\: V(x,y) .
\end{equation}
In practice, one gains access to the operator anomalous dimensions by renormalizing the partonic matrix elements of the operators in Eq.(\ref{eq:operators}). To do this, one has to choose a basis for the additional total-derivative operators. We focus in this work on two convenient options. The first, which we will call the Gegenbauer basis, is standard when one starts from conformal symmetry considerations and expresses the operators in terms of Gegenbauer polynomials~\cite{Braun:2017cih,Efremov:1978rn,Belitsky:1998gc}. The second basis to be discussed, denoted as the derivative basis, identifies operators by counting powers of derivatives and is commonly used when comparing non-perturbative lattice computations to the corresponding perturbative ones in the continuum, see e.g.~\cite{Gracey:2009da,Gockeler:2004wp}.

\subsection{The Gegenbauer basis}
Local operators in the Gegenbauer basis are defined in terms of Gegenbauer polynomials as follows~\cite{Efremov:1978rn,Belitsky:1998gc}
\begin{equation}
\label{eq:Gbasis}
    \mathcal{O}_{N,k}^{\mathcal{G}} = (\Delta \cdot \partial)^k \overline{\psi}(x)  (\Delta\cdot\Gamma) C_N^{3/2}\Bigg(\frac{\cev D \cdot \Delta-\Delta \cdot \Vec{D}}{\cev{\partial} \cdot \Delta+\Delta \cdot \Vec{\partial}}\Bigg)\psi(x).
\end{equation}
Here $\Delta$ represents an arbitrary lightlike vector, $\Delta^2=0$, which is used to select the leading-twist contributions. The Gegenbauer polynomials can be written as~\cite{olver10}
\begin{equation}
\label{eq:gegPol}
    C_N^{3/2}(z) = \frac{1}{2N!}\sum_{l=0}^{N}(-1)^l\binom{N}{l}\frac{(N+l+2)!}{(l+1)!}\Big(\frac{1}{2}-\frac{z}{2}\Big)^l.
\end{equation}
Quantities written in the Gegenbauer basis get a $\mathcal{G}$ superscript. $k$ represents the total number of derivatives and $k \geq N$. When expanding the Gegenbauer polynomial with the differential operators, the local operators can be rewritten as
\begin{equation}
    \mathcal{O}_{N,k}^{\mathcal{G}} = \frac{1}{2N!}\sum_{l=0}^{N}(-1)^l\binom{N}{l}\frac{(N+l+2)!}{(l+1)!} \sum_{j=0}^{k-l}\binom{k-l}{j}\overline{\psi}(x) (\Delta\cdot\Gamma) (\cev{D} \cdot \Delta)^{k-l-j}(\Delta \cdot \Vec{D})^{l+j}\psi(x).
\end{equation}
The renormalized operators, $[\mathcal{O}_{N,k}^{\mathcal{G}}]$, obey the following evolution equation
\begin{equation}
    \label{eq:evolG}
\Big(\mu^2 \partial_{\mu^2} + \beta(a_s)\partial_{a_s}\Big)[\mathcal{O}_{N,k}^{\mathcal{G}}] = \sum_{j=0}^{N}\gamma_{N,j}^{\mathcal{G}}[\mathcal{O}_{j,k}^{\mathcal{G}}]
\end{equation}
with $\beta(a_s)$ the standard QCD beta-function
\begin{equation}
     \frac{\text{d}\:a_s}{\text{d}\ln \mu^2} = \beta(a_s) =-a_s(\epsilon+\beta_0 a_s +\beta_1 a_s^2 +\beta_2 a_s^3 + \dots)
\end{equation}
and $\beta_0 = (11/3) C_A - (2/3) n_f$. The ADM in the Gegenbauer basis is triangular, i.e. $\gamma_{N,k}^{\mathcal{G}}=0$ if $k>N$. The diagonal elements, $\gamma_{N,N}$, simply correspond to the forward anomalous dimensions appearing in the description of inclusive scattering processes. As these do not depend on the basis chosen for the total-derivative operators, we can omit the superscript $\mathcal{G}$. Finally, when working in this basis one only considers CP-even operators. This forces the anomalous dimensions to vanish when $N-k$ is odd.

\subsection{The derivative basis}
Another convenient basis for the quark operators is
\begin{equation}
\label{eq:derBasis}
    \mathcal{O}_{{ p},{ q},{r}}^{\mathcal{D}} = (\Delta \cdot \partial)^{{ p}} \Big\{(\Delta \cdot D)^{{ q}}\overline{\psi}\, (\Delta \cdot\Gamma)
    (\Delta \cdot D)^{{r}}\psi\Big\},
\end{equation}
see e.g.~\cite{Gracey:2011zn, Gracey:2011zg}. Once again we introduced an arbitrary lightlike vector $\Delta$ to select the leading-twist contributions. The renormalization of the operators of interest in this basis is written as
\begin{equation}
    \mathcal{O}_{k,0,N}^{\mathcal{D}} = \sum\limits_{j=0}^{N}\, Z_{N,N-j}^{\D}\, [\mathcal{O}_{k+j,0,N-j}^{\mathcal{D}}]
\end{equation}
where the superscript $\mathcal{D}$ now represents the operators being written in the derivative basis. The operator anomalous dimensions are defined in terms of the renormalization constants in the standard way,
\begin{equation}
    \gamma_{N,k}^{\mathcal{D}}=-(Z_{N,j}^{\mathcal{D}})^{-1}\frac{\text{d}\:Z_{j,k}^{\mathcal{D}}}{\text{d}\ln \mu^2}.
\end{equation}
As was the case in the Gegenbauer basis, the derivative ADM is triangular, $\gamma_{N,k}^{\mathcal{D}}=0$ if $k>N$, and its diagonal elements are the basis-independent forward anomalous dimensions $\gamma_{N,N}$. \newline

It was shown in~\cite{Moch:2021cdq} that, when we assume the quarks to be massless, the anomalous dimensions in the derivative basis obey the following consistency relation
\begin{equation}
\label{mainConj}
    \gamma_{N,k}^{\mathcal{D}} \,=\, 
    \binom{N}{k}\sum_{j=0}^{N-k}(-1)^j \binom{N-k}{j}\gamma_{j+k,j+k} 
    + \sum_{j=k}^N (-1)^k \binom{j}{k} \sum_{l=j+1}^N (-1)^l \binom{N}{l} \gamma_{l,j}^{\mathcal{D}}
    \, .
\end{equation}
One can use this relation to reconstruct the off-diagonal part of the ADM when the diagonal elements and the last column, $\gamma_{N,0}^{\mathcal{D}}$, are known. The latter serves as a boundary condition for the consistency relation and, assuming one uses $D=4-2\eps$ dimensional regularization, can be related to the $1/\eps$-pole of the bare operator matrix elements, i.e. it can be determined through a Feynman diagram computation using standard computer algebra techniques. More details can be found in~\cite{Moch:2021cdq}. This method was used to compute the leading-$n_f$ anomalous dimensions up to order $a_s^5$ for the Wilson operators and to order $a_s^4$ for the transversity ones. Beyond the large-$n_f$ limit the consistency relation was also used to compute the order $a_s^2$ anomalous dimensions for the Wilson operators in the leading-color limit. Finally, using an implicit relation between the Gegenbauer basis and the derivative one,
\begin{equation}
\label{eq:impBasis}
    \sum\limits_{j=0}^{N}\, (-1)^j\frac{(j+2)!}{j!}\, \gamma_{N,j}^{\mathcal{G}} \,=\,
    \frac{1}{N!}\, \sum\limits_{j=0}^{N}\,
    (-1)^j\binom{N}{j}\frac{(N+j+2)!}{(j+1)!}\, \sum\limits_{l=0}^{j}\, \gamma_{j,l}^{\mathcal{D}}
    \, ,
\end{equation}
the results in the derivative basis were used to derive the leading-$n_f$ anomalous dimensions in the Gegenbauer basis to order $a_s^4$, both for the Wilson and the transversity operators. See~\cite{Moch:2021cdq,VanThurenhout:2022nmx} for more details on these computations.\newline

The relation between the Gegenbauer and derivative bases in Eq.(\ref{eq:impBasis}) is an implicit one, in the sense that it only relates sums of anomalous dimensions in one basis to sums of anomalous dimensions in the other. In the next section, we will derive a more explicit way to rotate between both bases. As we will see in the sections that follow, this will allow us to use the properties of the anomalous dimensions in both bases to our advantage.

\section{Construction of the similarity transformation matrix}
\label{sec:trans}
In this section, we will derive the explicit form of the rotation matrix to transform between the Gegenbauer basis and the derivative one. Specifically we are after an explicit representation of the matrix $\hat{R}$ where\footnote{Here and in the following, matrices are always denoted with a hat.}
\begin{equation}
\label{eq:basisTrans}
    \hat{\gamma}_{N}^{\mathcal{D}} = \hat{R}_{N}^{-1}\hat{\gamma}_{N}^{\mathcal{G}}\hat{R}_N.
\end{equation}
The rotation matrix $\hat{R}$ can be found by comparing the derivative structure of both bases. To do so, it is useful to consider the local operators as being derived from some generic non-local operator on the lightcone. Suppose we have some fundamental field $\phi(x)$ from which we construct a non-local operator $\phi(z_1)\phi(z_2)$. On the lightcone, one can apply an operator product expansion to construct local operators of the type~\cite{Braun:2003rp}
\begin{equation}
    \mathcal{O}_N(0) = \mathcal{P}_N(\partial_1,\partial_2)\phi(z_1)\phi(z_2)\big\vert_{z_{1,2}\rightarrow 0}
\end{equation}
where
\begin{equation}
    \partial_i \equiv \frac{\partial}{\partial_{z_i}}.
\end{equation}
$\mathcal{P}_N(z_1,z_2)$ represents the characteristic polynomial and has degree $N$. For example, the local operators in the Gegenbauer basis, cf.~Eq.(\ref{eq:Gbasis}), can be written as~\cite{Braun:2017cih}
\begin{equation}
\label{eq:localG}
    \mathcal{O}_{N,k}^{\mathcal{G}} = (\partial_1+\partial_2)^k C_N^{3/2}\Bigg(\frac{\partial_1-\partial_2}{\partial_1+\partial_2}\Bigg)\mathcal{O}(z_1,z_2)\big\vert_{z_{1,2}\rightarrow 0}
\end{equation}
with
\begin{equation}
    \mathcal{O}(z_1,z_2) = \Bar{\psi}(z_1 n)\slashed n \psi(z_2 n).
\end{equation}
The corresponding characteristic polynomial is
\begin{equation}
    \mathcal{P}_{N,k}^{\G}(z_1,z_2) = (z_1+z_2)^k C_{N}^{3/2}\Bigg(\frac{z_1-z_2}{z_1+z_2}\Bigg).
\end{equation}
This polynomial is homogeneous in the sense that
\begin{equation}
    (z_1\partial_1+x_2\partial_2-k)\mathcal{P}_{N,k}^{\G}(z_1,z_2) = 0.
\end{equation}
Here $k \geq N$ represents the total number of derivatives. We now derive the corresponding characteristic polynomial for the operators in the derivative basis. For this, we start from the fact that the non-local operators act as generating functions for local ones~\cite{Braun:2017cih}
\begin{equation}
    \mathcal{O}(z_1,z_2) = \sum_{k,m}\frac{z_1^k z_2^m}{k!m!}\mathcal{O}_{0,k,m}^{\mathcal{D}}.
\end{equation}
Note that the local operators appearing in the right-hand side are already written in the derivative basis. Using now the identity~\cite{VanThurenhout:2022hgd}
\begin{equation}
\label{eq:transBasis}
    \mathcal{O}_{0,N-k,k}^{\mathcal{D}} = (-1)^k\sum_{j=0}^{k}(-1)^j\binom{k}{j}\mathcal{O}_{j,N-j,0}^{\mathcal{D}}
\end{equation}
this can be rewritten in terms of operators with covariant derivatives acting only on the $\Bar{\psi}$ field
\begin{equation}
\label{eq:dExp}
    \mathcal{O}(z_1,z_2) = \sum_{k=0}^{N}\sum_{j=0}^{k}(-1)^{j+k}\binom{k}{j}\frac{z_1^{N-k}z_2^{k}}{k!(N-k)!}\mathcal{O}_{j,N-j,0}^{\mathcal{D}}.
\end{equation}
From this relation, the following two identities immediately follow
\begin{align}
    &\partial_1^N\mathcal{O}(z_1,z_2)\big\vert_{z_{1,2}\rightarrow 0} = \mathcal{O}_{0,N,0}^{\D}, \\ &\partial_2^N\mathcal{O}(z_1,z_2)\big\vert_{z_{1,2}\rightarrow 0} = \sum_{i=0}^{N}(-1)^{i}\binom{N}{i}\mathcal{O}_{N-i,i,0}^{\D}.
\end{align}
To determine the characteristic polynomial, we first consider the spin-two case, for which we only have two local operators $\{\mathcal{O}_{1,0,0}^{\mathcal{D}},\mathcal{O}_{0,1,0}^{\D}\}$. Using the identities just presented we find
\begin{align}
    &\partial_1 \Op(z_1,z_2)\big\vert_{z_{1,2}\rightarrow 0} = \Op_{0,1,0}^{\D},\\
    &\partial_2 \Op(z_1,z_2)\big\vert_{z_{1,2}\rightarrow 0} = \Op_{1,0,0}^{\D}-\Op_{0,1,0}^{\D}
\end{align}
and hence, upon inversion,
\begin{align}
    &\Op_{1,0,0}^{\D} = (\partial_1+\partial_2)\Op(z_1,z_2)\big\vert_{z_{1,2}\rightarrow 0},\\
    &\Op_{0,1,0}^{\D} = \partial_1\Op(z_1,z_2)\big\vert_{z_{1,2}\rightarrow 0}.
\end{align}
This implies that
\begin{align}
    &\mathcal{P}_{1,0}^{\D}(z_1,z_2) = z_1+z_2,\\
    &\mathcal{P}_{0,1}^{\D}(z_1,z_2) = z_1.
\end{align}
These polynomials are homogeneous with
\begin{equation}
    (z_1\partial_1+z_2\partial_2-N-k)\mathcal{P}_{N,k}^{\D}(z_1,z_2) = 0.
\end{equation}
Here $N,k\in\{0,1\}$ and $N+k=1$ represents the total number of derivatives. The above example can easily be generalized to higher-spin operators and we find
\begin{equation}
    \mathcal{P}_{N,k}^{\D}(z_1,z_2) = z_1^{k}(z_1+z_2)^N.
\end{equation}
This is a homogeneous polynomial of degree $N+k$, which is the total number of derivatives,
\begin{equation}
    (z_1\partial_1+z_2\partial_2-N-k)\mathcal{P}_{N,k}^{\D}(z_1,z_2) = 0.
\end{equation}
Hence we can conclude that the local operators in the derivative basis can be derived from the non-local one as
\begin{equation}
    \Op_{N,k,0}^{\D} = \partial_1^{k}(\partial_1+\partial_2)^{N}\Op(z_1,z_2)\big\vert_{z_{1,2}\rightarrow 0}.
\end{equation}
The equivalent relation for the operators with covariant derivatives acting on the $\psi$ field reads
\begin{equation}
\label{eq:localD}
    \Op_{N,0,k}^{\D} = \partial_2^{k}(\partial_1+\partial_2)^{N}\Op(z_1,z_2)\big\vert_{z_{1,2}\rightarrow 0}.
\end{equation}
Note that, in the limit of massless quarks, these operators have the same renormalization factors and hence the same anomalous dimensions. So, at least for our purposes, they are equivalent and we will focus on the latter operators in what follows. At this point we have all necessary ingredients to derive the rotation matrix between the derivative and Gegenbauer bases. Comparing Eqs.(\ref{eq:localG}) and (\ref{eq:localD}) we see that, in order to write one in terms of the other, we can rewrite the Gegenbauer polynomial in Eq.(\ref{eq:localG}) as
\begin{equation}
    C_N^{3/2}\Bigg(\frac{\partial_1-\partial_2}{\partial_1+\partial_2}\Bigg) = C_N^{3/2}\Bigg(1-\frac{2\partial_2}{\partial_1+\partial_2}\Bigg) \equiv C_N^{3/2}(1-2z)
\end{equation}
and compute the expansion
\begin{equation}
\label{eq:expGeg}
    C_{N}^{3/2}(1-2z) = \sum_{k=0}^{N}c_{N,k}z^k.
\end{equation}
The elements of the sought-after rotation matrix, $R_{N,k}$, will then simply coincide with the expansion coefficients $c_{N,k}$. From the definition of the Gegenbauer polynomials, cf.~Eq.(\ref{eq:gegPol}), it is easy to see that
\begin{equation}
    C_{N}^{3/2}(1-2z) = \frac{1}{2N!}\sum_{k=0}^{N}(-1)^k\binom{N}{k}\frac{(N+k+2)!}{(k+1)!}z^k
\end{equation}
and hence we finally find
\begin{equation}
\label{eq:R}
    R_{N,k} = \frac{1}{2N!}(-1)^k\binom{N}{k}\frac{(N+k+2)!}{(k+1)!}.
\end{equation}
To write down the complete basis transformation in Eq.(\ref{eq:basisTrans}) we now need to compute the elements of the inverse rotation matrix, $R^{-1}_{N,k}\equiv\Tilde{R}_{N,k}$. For this we start from Eq.(\ref{eq:expGeg}) to write
\begin{equation}
\label{eq:zn}
    z^N = \sum_{j=0}^{N}\Tilde{R}_{N,j}\:C_{j}^{3/2}(1-2z).
\end{equation}
Using that the Gegenbauer polynomials are orthogonal~\cite{gradshteyn2007}
\begin{equation}
\label{eq:GegOrth}
    \int_{0}^{1}\text{d}z\:C_{N}^{3/2}(1-2z)C_{k}^{3/2}(1-2z)[4z(1-z)] = \frac{(N+1)(N+2)}{3+2N}\delta_{N,k}
\end{equation}
we can multiply both sides of Eq.(\ref{eq:zn}) with $4z(1-z)C_{k}^{3/2}(1-2z)$ and integrate over $z$ to find
\begin{equation}
    \Tilde{R}_{N,k} = \frac{4(3+2k)}{(k+1)(k+2)}\int_{0}^{1}\text{d}z\:z^{N+1}(1-z)C_{k}^{3/2}(1-2z).
\end{equation}
To compute the integral on the right-hand side, one can use the following identity for the Gegenbauer polynomials~\cite{gradshteyn2007}
\begin{equation}
\label{eq:GegDer}
    C_{N}^{3/2}(t) = \frac{(-1)^N (N+2)}{2^{N+1}N!}\frac{1}{1-t^2}\frac{\text{d}^N}{\text{d}t^N}(1-t^2)^{N+1}
\end{equation}
to write
\begin{equation}
    \Tilde{R}_{N,k} = \frac{3+2k}{2^{2k+1}(k+1)k!}\int_{0}^{1}\text{d}z\:z^{N}\frac{\text{d}^k}{\text{d}z^k}[4z(1-z)]^{k+1}.
\end{equation}
The remaining integral can be computed using integration-by-parts and we finally find
\begin{equation}
\label{eq:RTilde}
    \Tilde{R}_{N,k} = \frac{2(-1)^k(3+2k)\Gamma(N+1)\Gamma(N+2)}{\Gamma(N-k+1)\Gamma(N+k+4)} = \frac{2(-1)^k(3+2k)k!(N+1)!}{(N+k+3)!}\binom{N}{k}
\end{equation}
with $\Gamma(N)$ the standard gamma function. With the elements of both the rotation matrix $\hat{R}$ and its inverse $\hat{R}^{-1}$ computed, the transformation between the Gegenbauer basis and the derivative one
in Eq.(\ref{eq:basisTrans}) can be implemented. In component form we can now write it as
\begin{align}
    \gamma_{N,k}^{\D} &= \sum_{l=0}^{N}\sum_{j=k}^{l}\Tilde{R}_{N,l}R_{j,k}\gamma_{l,j}^{\G}\\
    &=\frac{(-1)^k(N+1)!}{(k+1)!}\sum_{l=k}^{N}(-1)^l\binom{N}{l}\frac{l!(3+2l)}{(N+l+3)!}\sum_{j=k}^{l}\binom{j}{k}\frac{(j+k+2)!}{j!}\gamma_{l,j}^{\G}.
\end{align}
In the second line we changed the lower limit of $l$ from $0$ to $k$, as for $l<k$ the inner sum would be empty. Inverting the relation we have
\begin{equation}
\label{eq:basisTransIn}
    \hat{\gamma}^{\G} = \hat{R}\hat{\gamma}^{\D}\hat{R}^{-1}
\end{equation}
and in component form
\begin{align}
\label{eq:DtoG}
    \gamma_{N,k}^{\G} &= \sum_{l=0}^{N}\sum_{j=k}^{l}R_{N,l}\Tilde{R}_{j,k}\gamma_{l,j}^{\D}\\
    &=(-1)^k\frac{k!}{N!}(3+2k)\sum_{l=k}^{N}(-1)^{l}\binom{N}{l}\frac{(N+l+2)!}{(l+1)!}\sum_{j=k}^{l}\binom{j}{k}\frac{(j+1)!}{(j+k+3)!}\gamma_{l,j}^{\D}.
\end{align}
We again shifted the lower limit of $l$ from $0$ to $k$ in the second line. We emphasize that, while the transformation rules have been explicitly given for the anomalous dimensions, they are generic, i.e. Eqs.(\ref{eq:basisTrans}) and (\ref{eq:basisTransIn}) can be used to relate any quantity in the Gegenbauer basis to the corresponding one in the derivative basis and vice versa. Moreover, as the derivation was independent of the Dirac structure of the local operators under consideration, the derived basis transformation formulae are valid for generic leading-twist quark operators.\newline

When considering the renormalization of flavor-singlet operators, one also has to take into account the mixing between quark and gluon operators with the same quantum numbers. We now set up the basis transformation between the Gegenbauer and derivative bases for the latter. In the derivative basis, the flavor-singlet Wilson quark operators are written as
\begin{align}
\label{eq:singQ}
    \mathcal{O}_{k,0,N}^{q,\D} &= (\Delta\cdot\partial)^k\overline{\psi}\slashed\Delta (\Delta\cdot D)^N\psi
\end{align}
while the gluon operators are of the form
\begin{equation}
\label{eq:singG}
    \mathcal{O}_{k,0,N-1}^{g,\D} = (\Delta\cdot\partial)^k F_{\mu\Delta} (\Delta\cdot D)^{N-1}F_{\Delta}^{\:\:\:\:\mu}.
\end{equation}
Here $\Delta$ is an arbitrary lightlike vector and $F_{\mu\Delta}$ is understood to mean
\begin{equation}
    F_{\mu\Delta} = F_{\mu\nu}\Delta^{\nu}
\end{equation}
with $F_{\mu\nu}$ the standard gluon field strength of QCD. Meanwhile in the Gegenbauer basis we have~\cite{Braun:2022byg}
\begin{equation}
    \mathcal{O}_{N,k}^{g,\G} = 6(\Delta\cdot\partial)^{k-1}F_{\mu\Delta}C_{N-1}^{5/2}\Bigg(\frac{\cev D \cdot \Delta-\Delta \cdot \Vec{D}}{\cev{\partial} \cdot \Delta+\Delta \cdot \Vec{\partial}}\Bigg)F^{\mu\Delta}.
\end{equation}
Note the shifts in $N$ and $k$ here. These originate from the fact that no spin-one twist-two gluon operator exists and from the condition $k\geq N$ in this operator basis. The Gegenbauer polynomials $C_{N}^{5/2}(z)$ can be written as~\cite{olver10}
\begin{equation}
\label{eq:geg52}
    C_{N}^{5/2}(z) =  \frac{1}{12\, N!}\, 
  \sum\limits_{l=0}^{N}\, (-1)^l\, 
  \binom{N}{l}\,
  \frac{(N+l+4)!}{(l+2)!}\, 
  \left(\frac{1}{2}-\frac{1}{2}z\right)^l
  \, .
\end{equation}
Expanding the Gegenbauer polynomial with the differential operator we can rewrite the local gluon operators as
\begin{align}
    \Op_{N,k}^{g,\G} = &\frac{1}{2(N-1)!}\sum_{l=0}^{N-1}(-1)^l \binom{N-1}{l}\frac{(N+l+3)!}{(l+2)!}\nonumber\\&\times\sum_{j=0}^{k-1-l}\binom{k-1-l}{j}F_{\mu\Delta}(\cev D\cdot\Delta)^{k-l-j-1}(\Delta\cdot\Vec{D})^{l+j}F^{\mu\Delta}.
\end{align}
We now compute the gluon rotation matrix, denoted by $\hat{G}$, to transform between the Gegenbauer and derivative bases,
\begin{equation}
\label{eq:gluon1}
    \hat{\gamma}^{g,\D} = \hat{G}^{-1}\hat{\gamma}^{g,\G}\hat{G}.
\end{equation}
As before, the elements of the rotation matrix $\hat{G}$ correspond to the expansion coefficients of the Gegenbauer polynomial in $1-2z$. The only difference with the computation above is that now we need to consider $C_{N-1}^{5/2}(1-2z)$. From Eq.(\ref{eq:geg52}) we see that
\begin{equation}
    6\:C_{N-1}^{5/2}(1-2z) = \frac{1}{2\, (N-1)!}\, 
  \sum\limits_{l=0}^{N-1}\, (-1)^l\, 
  \binom{N-1}{l}\,
  \frac{(N+l+3)!}{(l+2)!}\, 
  z^l
\end{equation}
and hence the elements of the rotation matrix $\hat{G}$ are
\begin{equation}
\label{eq:gluon2}
    G_{N,k} = -\frac{1}{2(N-1)!}(-1)^k\binom{N-1}{k-1}\frac{(N+k+2)!}{(k+1)!}
\end{equation}
in which an additional $k\rightarrow k-1$ shift was introduced to incorporate the fact that, in the derivative basis, $k$ should be larger than zero\footnote{This is because $k=0$ would represent mixing into a spin-one operator and its derivatives.}. For the derivation of the elements of the inverse transformation matrix, $\hat{G}^{-1}\equiv\Tilde{G}$, we follow the same procedure as above. To do so, we need the generalizations of Eqs.(\ref{eq:GegOrth}) and (\ref{eq:GegDer}), which are~\cite{gradshteyn2007}
\begin{equation}
    \int_{0}^{1}\text{d}z\:C_{N}^{5/2}(1-2z)C_{k}^{5/2}(1-2z)[4z(1-z)]^2 = \frac{(N+1)(N+2)(N+3)(N+4)}{9(5+2N)}\delta_{N,k}
\end{equation}
and
\begin{equation}
    C_{N}^{5/2}(t) = \frac{(-1)^N (N+3)(N+4)}{12\cdot2^{N}N!}\frac{1}{(1-t^2)^2}\frac{\text{d}^N}{\text{d}t^N}(1-t^2)^{N+2}.
\end{equation}
Following the same steps as before we then find
\begin{equation}
\label{eq:gluon3}
    \Tilde{G}_{N,k} = -\frac{2(-1)^{k}(3+2k)\Gamma(N)\Gamma(N+2)}{\Gamma(N-k+1)\Gamma(N+k+4)} = -\frac{2(-1)^{k}(3+2k)k!(N+1)!}{N(N+k+3)!}\binom{N}{k}.
\end{equation}

\subsection{Application: One-loop anomalous dimensions}
As a simple application of the basis transformation presented above, we consider the operator anomalous dimensions at one loop. At this order in perturbation theory, the ADM in the Gegenbauer basis is diagonal due to the exact conformal symmetry of QCD at the critical coupling at leading order, see e.g.~\cite{Efremov:1978rn,Makeenko:1980bh}. This means that in the transformation rule
\begin{align}
    \gamma_{N,k}^{\D} 
    &=\frac{(-1)^k(N+1)!}{(k+1)!}\sum_{l=k}^{N}(-1)^l\binom{N}{l}\frac{l!(3+2l)}{(N+l+3)!}\sum_{j=k}^{l}\binom{j}{k}\frac{(j+k+2)!}{j!}\gamma_{l,j}^{\G}
\end{align}
the inner sum collapses to just the diagonal term and we are left with a single sum over the forward anomalous dimensions
\begin{align}
        \gamma_{N,k}^{\D,(0)} &= \frac{(-1)^k N!(N+1)!}{(k+1)!}\sum_{l=k}^{N}(-1)^l\binom{l}{k}\frac{(3+2l)(l+k+2)!}{l!(N-l)!(N+l+3)!}\gamma_{l,l}^{(0)}.
    \end{align}
Note that this relation is valid for any leading-twist quark operator. Setting $k=N-1$ we recover
\begin{equation}
\label{eq:NTD}
    \gamma_{N,N-1}^{\D,(0)} = \frac{N}{2}(\gamma_{N-1,N-1}^{(0)}-\gamma_{N,N}^{(0)})
\end{equation}
which we previously derived in~\cite{Moch:2021cdq} using our consistency relation, cf.~Eq.(\ref{mainConj}). However, Eq.(\ref{mainConj}) actually implies that Eq.(\ref{eq:NTD}) should be valid to all orders in perturbation theory. The representation of the other matrix elements of the ADM in terms of the forward anomalous dimensions is new, but only valid at the one-loop level.
Substituting the explicit values for the diagonal elements~\cite{Gross:1973ju,Floratos:1977au,Artru:1989zv,Shifman:1980dk,Baldracchini:1981,Blumlein:2001ca}, we find
\begin{align}
\label{eq:W1l}
    \gamma_{N,k}^{\D,(0)} &= \frac{(-1)^k N!(N+1)!}{(k+1)!}C_F\sum_{l=k}^{N}(-1)^l\binom{l}{k}\frac{(3+2l)(l+k+2)!}{l!(N-l)!(N+l+3)!}\Big(4S_1(l)+\frac{2}{l+1}+\frac{2}{l+2}-3\Big)\nonumber\\&=2\:C_F\Big(\frac{1}{N+2}-\frac{1}{N-k}\Big)
\end{align}
for the Wilson operators and
\begin{align}
\label{eq:T1l}
    \gamma_{N,k}^{T,\D,(0)} &= \frac{(-1)^k N!(N+1)!}{(k+1)!}C_F\sum_{l=k}^{N}(-1)^l\binom{l}{k}\frac{(3+2l)(l+k+2)!}{l!(N-l)!(N+l+3)!}\Big(4S_1(l)+\frac{4}{l+1}-3\Big)\nonumber\\&=2\:C_F\Big(\frac{1}{N+1}-\frac{1}{N-k}\Big)
\end{align}
for the transversity ones. Here $C_F=\frac{n_c^2-1}{2n_c}$ represents the quadratic Casimir operator of the fundamental representation of $SU(n_c)$. The harmonic sums are recursively defined as~\cite{Vermaseren:1998uu,Blumlein:1998if}
\begin{eqnarray}
\label{eq:Hsum}
  S_{\pm m}(N) &=& \: \: \sum_{i=1}^{N}\; (\pm 1)^i \, i^{\, -m}
  \, , 
  \nonumber\\
  S_{\pm m_1^{},\,m_2^{},\,\ldots,\,m_d}(N) &=& \: \: \sum_{i=1}^{N}\:
  (\pm 1)^{i} \; i^{\, -m_1^{}}\; S_{m_2^{},\,\ldots,\,m_d}(i)
\, .
\end{eqnarray}
Both expressions agree with previous calculations~\cite{Artru:1989zv,Shifman:1980dk,Baldracchini:1981,Blumlein:2001ca,Moch:2021cdq,VanThurenhout:2022nmx}. The sums in Eqs.(\ref{eq:W1l}) and (\ref{eq:T1l}) were evaluated using the \textit{M{\small ATHEMATICA}} package \textit{S{\small IGMA}}~\cite{Schneider2007,Schneider:2013uan}.

\subsection{Structure of the anomalous dimensions beyond one loop accuracy}
Beyond the one-loop level, the anomalous dimensions in the $\D$-basis receive corrections from the off-diagonal elements in the Gegenbauer basis. Schematically, in terms of the rotation matrix $R$, we have
\begin{equation}
\label{eq:approxADM}
    \gamma_{N,k}^{\D} = \sum_{l=0}^{N}\Tilde{R}_{N,l}\gamma_{l,l}R_{l,k}+\sum_{l=0}^{N}\Tilde{R}_{N,l}\sum_{j=0}^{l-1}\gamma_{l,j}^{\G}R_{j,k}.
\end{equation}
Here the first sum only gets contributions from the forward anomalous dimensions, while the second sum only gets contributions from the off-diagonal elements in the Gegenbauer basis. However, because of the structure of the ADM (which is determined by conformal symmetry) in the latter, these off-diagonal elements are completely determined by lower-order quantities. For example, at the two-loop level we have\footnote{Because of different conventions in the definition of the anomalous dimensions, there are additional factors of 2 here as compared to the expressions in~\cite{Braun:2017cih}.} \cite{Braun:2017cih}
\begin{equation}
\label{eq:geg2L}
    \gamma_{N,k}^{\mathcal{G}, (1)} \,=\, \delta_{N,k}\gamma_{k,k}^{(1)}
     -\frac{\gamma_{N,N}^{(0)}-\gamma_{k,k}^{(0)}}{a(N,k)}\Bigg\{-2(2k+3)\Big(\beta_0+\gamma_{k,k}^{(0)}\Big)\vartheta_{N,k}+2\Delta_{N,k}^{\G,(0)}\Bigg\}
    \, 
\end{equation}
with 
\begin{equation}
\label{eq:defa}
     a(N,k) = (N-k)(N+k+3).
\end{equation}
The discrete step-function is defined as
\begin{equation}
\vartheta_{N,k} \,\equiv\,
    \begin{cases}
     1 \quad \text{if\:} N-k > 0 \text{\:and even} \\  0 \quad \text{else}
    \end{cases}
\end{equation}
and appears because of CP-symmetry in the Gegenbauer basis. The quantity $\Delta$ is the so-called \textit{conformal anomaly}. In conformal schemes, it signals the breakdown of exact conformal symmetry of QCD beyond leading order. The conformal anomaly can be computed perturbatively,
\begin{equation}
    \Delta_{N,k}^{\G}(a_s) = a_s\Delta_{N,k}^{\G,(0)}+a_s^2\Delta_{N,k}^{\G,(1)}+\dots \:.
\end{equation}
Currently it is known to order $a_s^2$, although a closed-form expression for $\Delta_{N,k}^{(1)}$ is not available at this time~\cite{Braun:2017cih,Braun:2016qlg}.\newline So we see that, at two loop accuracy, the off-diagonal elements of the Gegenbauer ADM are completely determined by the one-loop forward anomalous dimensions, one-loop QCD beta-function and one-loop conformal anomaly. In turn, this implies that only these one-loop quantities appear in the second sum in Eq.(\ref{eq:approxADM}). Moreover, as generically the anomalous dimensions in the Gegenbauer basis at $L$ loops require knowledge of the conformal anomaly to $(L-1)$-loop accuracy~\cite{Mueller:1991gd}, it follows that the second sum in Eq.(\ref{eq:approxADM}) always involves only quantities at lower orders in perturbation theory. To summarize, we find that
\begin{itemize}
    \item at the one-loop level, the off-diagonal elements of the ADM in the $\D$-basis are completely determined by the forward anomalous dimensions;
    \item at higher orders in perturbation theory, the off-diagonal elements of the $\D$-basis ADM are determined by a single sum over forward anomalous dimensions and a double sum involving only lower-order quantities.
\end{itemize}
Finally we note that, while we focused now on the quark operators, the discussion above is also valid for gluon operators, see e.g. Sec.\ref{sec:gluon} below for an application.

\subsection{Consistency relation from CP-symmetry}
Consider the inverse basis transformation,
\begin{equation}
\label{eq:invTrans}
    \gamma_{N,k}^{\G} = R_{N,l}\gamma_{l,j}^{\D}\Tilde{R}_{j,k}.
\end{equation}
As mentioned above, the operators in the Gegenbauer basis are CP-even. At the level of the anomalous dimensions, this means that $\gamma_{N,k}^{\G}=0$ whenever $N-k$ is odd. Substituting into Eq.(\ref{eq:invTrans}) we then find a tower of relations for the anomalous dimensions in the derivative basis
\begin{equation}
    \sum_{l=k}^{N}(-1)^{l}\binom{N}{l}\frac{(N+l+2)!}{(l+1)!}\sum_{j=k}^{l}\binom{j}{k}\frac{(j+1)!}{(j+k+3)!}\gamma_{l,j}^{\D} = 0 \: \: \: \: (N-k \text{\:odd}).
\end{equation}
Note that these relations are valid to all orders of perturbation theory. Setting $k=N-1$ we reproduce
\begin{equation}
    \gamma_{N,N-1}^{\D} = \frac{N}{2}(\gamma_{N-1,N-1}-\gamma_{N,N})
\end{equation}
while for $k=N-3$ we find
\begin{align}
    &2\gamma_{N,N-3}^{\D}+\frac{1}{6}N(N-1)(N-2)(\gamma_{N,N}-\gamma_{N-3,N-3})+(N-2)\gamma_{N,N-2}^{\D}\nonumber\\&+\frac{1}{2}N(N-1)\gamma_{N-2,N-3}^{\D}+\frac{1}{2}(N-1)(N-2)\gamma_{N,N-1}^{\D}-N\gamma_{N-1,N-3}^{\D}=0.
\end{align}
Both of these expressions exactly match the corresponding ones derived from our consistency relation in Eq.(\ref{mainConj}), and the same holds for other $(N-\alpha)$-values with $\alpha$ odd. Hence, we can conclude that the physical origin of this consistency condition is tied to CP symmetry.

\section{All-order anomalous dimensions in the large-$n_f$ limit}
\label{sec:largeNF}
In this section we use our basis transformation to compute the anomalous dimensions in the Gegenbauer basis, in the large-$n_f$ limit, to all orders in perturbation theory. The corresponding computation in the derivative basis was performed in~\cite{VanThurenhout:2022hgd} and generalized previous calculations for the forward anomalous dimensions based on exact conformal symmetry at the Wilson-Fisher critical point~\cite{Gracey:1994nn,Gracey:2003mr}. For the sake of explicitness, we focus our attention on the spin-five ADM. Generalization to higher-spin operators is straightforward. After a brief review of the method, we present the anomalous dimensions in the Gegenbauer basis to sixth order in $a_s$. The corresponding all-order expressions, both in the derivative and the Gegenbauer bases, are collected in Appendix~\ref{sec:appA}.

\subsection{Review of the method}
As explained in~\cite{VanThurenhout:2022hgd}, to derive the all-order anomalous dimensions in the large-$n_f$ limit we start from the action of the evolution operator on the non-local lightcone operators
\begin{align}
    \begin{split}
    \label{eq:allorder}
        \gamma\mathcal{O}(z_1,z_2) =& \: \: \frac{\mu(\mu-1)}{2(\mu-2)(2\mu-1)}\eta\Bigg\{\int_{0}^{1}\text{d}\alpha\:\frac{\overline{\alpha}^{\mu-1}}{\alpha}(2[\mathcal{O}(z_1,z_2)]-[\mathcal{O}(z_{12}^{\alpha},z_2)]-[\mathcal{O}(z_1,z_{21}^{\alpha})])\\&-(\mu-\delta)^2\int_{0}^{1}\text{d}\alpha\int_{0}^{\overline{\alpha}}\text{d}\beta\:(1-\alpha-\beta)^{\mu-2}[\mathcal{O}(z_{12}^{\alpha},z_{21}^{\beta})]+\frac{\mu-1}{\mu}\mathcal[{O}(z_1,z_2)]\Bigg\}
    \end{split}
\end{align}
with $z_{12}^{\alpha} = z_1\overline{\alpha}+z_2\alpha$ and $\overline{\alpha} = 1-\alpha$. The Dirac structure is controlled by the parameter $\delta$, which is $\delta=1$ for Wilson operators and $\delta=2$ for transversity ones. The parameters $\mu$ and $\eta$ are defined as
\begin{equation}
    \mu = \frac{D}{2} = 2-\eps
\end{equation}
and
\begin{equation}
    \eta = \frac{1}{n_f}\frac{(\mu-2)(2\mu-1)\Gamma(2\mu)}{\Gamma^2(\mu)\Gamma(\mu+1)\Gamma(2-\mu)}
\end{equation}
respectively, in which $\eps$ is understood to be evaluated at the Wilson-Fisher fixed point,
\begin{equation}
    \eps\rightarrow\eps_{*} = -a_s\beta_0\biggr\vert_{n_f} = \frac{2}{3}n_fa_s.
\end{equation}
Expanding the non-local operators in Eq.(\ref{eq:allorder}) in terms of local ones, written in the derivative basis, we find
\begin{align}
    \gamma\Op(z_1,z_2) =& \: \: \frac{\mu(\mu-1)}{2(\mu-2)(2\mu-1)}\eta\sum_{k=0}^{N}\sum_{j=0}^{k}(-1)^{j+k}\binom{k}{j}\frac{[\Op_{j,N-j,0}^{\D}]}{k!(N-k)!}\Bigg\{\int_{0}^{1}\text{d}\alpha\:\frac{\overline{\alpha}^{\mu-1}}{\alpha}\big\{2z_1^{N-k}z_2^{k}\nonumber\\&-(z_1\overline{\alpha}+z_2\alpha)^{N-k}z_2^{k}-z_1^{N-k}(z_2\overline{\alpha}+z_1\alpha)^k\big\}\nonumber\\&-(\mu-\delta)^2\int_{0}^{1}\text{d}\alpha\int_{0}^{\overline{\alpha}}\text{d}\beta\:(1-\alpha-\beta)^{\mu-2}\big\{(z_1\overline{\alpha}+z_2\alpha)^{N-k}(z_2\overline{\beta}+z_1\beta)^{k}\big\}\nonumber\\&+\frac{\mu-1}{\mu}z_1^{N-k}z_2^{k}\Bigg\}.
\end{align}
Next, one evaluates this expression for the particular $N$-value in which one is interested and takes the $N$-th derivative with respect to $z_1$. Finally, after setting $z_{1,2}\rightarrow 0$ one obtains an expansion of the form
\begin{equation}
    \frac{\text{d}^{N}}{\text{d}z_1^{N}}\gamma\Op(z_1,z_2)\biggr\vert_{z_{1,2}\rightarrow\: 0} = \sum_{j=0}^{N}\gamma_{N,j}^{\D}[\Op_{N-j,j,0}^{\D}]
\end{equation}
from which one can read off the all-order expressions for the anomalous dimensions in the derivative basis.

\subsection{Spin-five Gegenbauer anomalous dimensions to six loops}
We now present the leading-$n_f$ Gegenbauer ADM for the spin-five operators to sixth order in the strong coupling. These were obtained by first computing the corresponding all-order results in the derivative basis as described above. Next we apply the basis transformation formula, Eq.(\ref{eq:basisTransIn}), to obtain the all-order results in the Gegenbauer basis. Finally, the latter are expanded up to $a_s^6$. Writing the spin-five Gegenbauer ADM as
\begin{equation}
    \begin{pmatrix}
        \gamma_{4,4} & 0 & \gamma_{4,2}^{\G} & 0 & \gamma_{4,0}^{\G}\\
        0 & \gamma_{3,3} & 0 & \gamma_{3,1}^{\G} & 0 \\
        0 & 0 & \gamma_{2,2} & 0 & \gamma_{2,0}^{\G} \\
        0 & 0 & 0 & \gamma_{1,1} & 0 \\
        0 & 0 & 0 & 0 & 0
    \end{pmatrix}
\end{equation}
we find the following expansions for the off-diagonal elements
\begin{align}
    \gamma_{4,2}^{\G} =& -\frac{133 }{135}a_s^2 n_f
+\frac{10339 }{12150}a_s^3 n_f^2
+\frac{394793 }{1093500}a_s^4 n_f^3
+\Bigg(\frac{17835391 }{98415000}
-\frac{2128  }{3645}\zeta_{3}\Bigg)a_s^5 n_f^4
\nonumber\\&+\Bigg(\frac{965662817 }{8857350000}
-\frac{2128  }{3645}\zeta_{4}
+\frac{82712  }{164025}\zeta_{3}\Bigg)a_s^6 n_f^5 + O(a_s^7),\\
\gamma_{4,0}^{\G} =& -\frac{13 }{15}a_s^2 n_f
+\frac{329 }{1350}a_s^3 n_f^2
+\frac{65923 }{121500}a_s^4 n_f^3
+\Bigg(\frac{4992701}{10935000}
-\frac{208}{405}  \zeta_{3}\Bigg)a_s^5 n_f^4
\nonumber\\&+\Bigg(\frac{300606787 }{984150000}
-\frac{208}{405}  \zeta_{4}
+\frac{2632 }{18225}\zeta_{3}\Bigg) a_s^6 n_f^5+O(a_s^7),\\
\gamma_{3,1}^{\G} =& -\frac{11 }{9}a_s^2 n_f
+\frac{793 }{810}a_s^3 n_f^2
+\frac{36491 }{72900}a_s^4 n_f^3
+\Bigg(\frac{1707517 }{6561000}
-\frac{176}{243}  \zeta_{3}\Bigg)a_s^5 n_f^4
\nonumber\\&+\Bigg(\frac{92204579 }{590490000}
-\frac{176}{243}  \zeta_{4}
+\frac{6344  }{10935}\zeta_{3}\Bigg)a_s^6 n_f^5+ O(a_s^7),\\
\gamma_{2,0}^{\G} =& -\frac{5 }{3}a_s^2 n_f
+\frac{59 }{54}a_s^3 n_f^2
+\frac{797 }{972}a_s^4 n_f^3
+\Bigg(\frac{8543 }{17496}
-\frac{80}{81}  \zeta_{3}\Bigg)a_s^5 n_f^4
\nonumber\\&+\Bigg(\frac{93005 }{314928}
-\frac{80}{81}  \zeta_{4}
+\frac{472}{729} \zeta_{3}\Bigg)a_s^6 n_f^5+ O(a_s^7).
\end{align}
The results up to order $a_s^4$ were computed before, see~\cite{Braun:2017cih,Moch:2021cdq}. The higher-order results are new. Note the appearance of the Riemann-zeta function
\begin{equation}
    \zeta_{N} = \sum_{j=0}^{\infty}\frac{1}{j^N}
\end{equation}
starting at the five-loop level. The above expressions are for the Wilson operators. Repeating the procedure for transversity operators we find
\begin{align}
    \gamma_{4,2}^{T,\G} =& -\frac{14 }{15}a_s^2 n_f
+\frac{511 }{675}a_s^3 n_f^2
+\frac{22057 }{60750}a_s^4 n_f^3
+\Bigg(\frac{1105559 }{5467500}
-\frac{224}{405}  \zeta_{3}\Bigg)a_s^5 n_f^4
\nonumber\\&+\Bigg(\frac{61019833 }{492075000}
-\frac{224}{405}  \zeta_{4}
+\frac{8176  }{18225}\zeta_{3}\Bigg)a_s^6 n_f^5+ O(a_s^7),\\
\gamma_{4,0}^{T,\G} =& -\frac{11 }{15}a_s^2 n_f
+\frac{253 }{1350}a_s^3 n_f^2
+\frac{45311 }{121500}a_s^4 n_f^3
+\Bigg(\frac{3788857 }{10935000}
-\frac{176}{405}  \zeta_{3}\Bigg)a_s^5 n_f^4
\nonumber\\&+\Bigg(\frac{288113159}{984150000}
-\frac{176}{405}  \zeta_{4}
+\frac{2024  }{18225}\zeta_{3}\Bigg) a_s^6 n_f^5+ O(a_s^7),\\
\gamma_{3,1}^{T,\G} =& -\frac{10 }{9}a_s^2 n_f
+\frac{65 }{81}a_s^3 n_f^2
+\frac{683 }{1458}a_s^4 n_f^3
+\Bigg(\frac{38581 }{131220}
-\frac{160}{243}  \zeta_{3}\Bigg)a_s^5 n_f^4
\nonumber\\&+\Bigg(\frac{2297147 }{11809800}
-\frac{160}{243}  \zeta_{4}
+\frac{1040  }{2187}\zeta_{3}\Bigg)a_s^6 n_f^5+ O(a_s^7),\\
\gamma_{2,0}^{T,\G} =& -\frac{4 6}{3}a_s^2 n_f
+\frac{20 }{27}a_s^3 n_f^2
+\frac{136 }{243}a_s^4 n_f^3
+\Bigg(\frac{938 }{2187}
-\frac{64}{81}  \zeta_{3}\Bigg)a_s^5 n_f^4
\nonumber\\&+\Bigg(\frac{7033 }{19683}
-\frac{64}{81}  \zeta_{4}
+\frac{320}{729}  \zeta_{3}\Bigg)a_s^6 n_f^5+ O(a_s^7).
\end{align}
The results up to order $a_s^4$ were computed before in~\cite{VanThurenhout:2022hgd}, while the terms of order $a_s^5$ and $a_s^6$ are new. 

\section{Two-loop anomalous dimensions beyond leading color}
In this section, we provide a way to compute the two-loop anomalous dimensions in the derivative basis from the corresponding results in the Gegenbauer one. In the derivative basis, low-$N$ results were computed for the Wilson operators in~\cite{Gracey:2009da} while general $(N,k)$ expressions for the large-$n_f$ and leading-color approximations were presented in~\cite{Moch:2021cdq}. The latter results were obtained by combining a computation of the operator matrix elements with our consistency relation, cf.~Eq.(\ref{mainConj}). Unfortunately, our current knowledge concerning the anomalous dimensions beyond the leading-color limit is limited. Let us decompose the two-loop anomalous dimensions as follows
\begin{equation}
\label{eq:colDecom}
    \gamma_{N,k}^{(1)} = \gamma_{N,k}^{(1)}\biggr\vert_{n_f}+\gamma_{N,k}^{(1)}\biggr\vert_{\text{LC}}+\gamma_{N,k}^{(1)}\biggr\vert_{\text{SLC}}.
\end{equation}
We will focus on the Wilson operators in what follows. The first term in Eq.(\ref{eq:colDecom}) represents the large-$n_f$ limit while the remaining terms represent the leading- and subleading-color terms respectively. In terms of the color and flavor factors we have
\begin{align}
    &\gamma_{N,k}^{(1)}\biggr\vert_{n_f}\sim n_f C_F, \\&
    \gamma_{N,k}^{(1)}\biggr\vert_{\text{LC}} \sim C_F^2, \\&
    \gamma_{N,k}^{(1)}\biggr\vert_{\text{SLC}} \sim C_F\Big(C_F-\frac{C_A}{2}\Big)
\end{align}
with $C_F=\frac{n_c^2-1}{2n_c}$ the quadratic Casimir of the fundamental representation of $SU(n_c)$ and $C_A=n_c$ the quadratic Casimir of the adjoint one.
For completeness, we repeat here the expressions of the first two limits, computed in~\cite{Moch:2021cdq}
\begin{align}
\gamma_{N,k}^{\mathcal{D}, (1)}\biggr\vert_{n_f} =& \: \: \frac{4}{3}n_f C_F \Bigg\{\Big(S_{1}(N)-S_{1}(k)\Big)  \Big( \frac{1}{N+2} - \frac{1}{N-k}  \Big)
       + \frac{5}{3}\frac{1}{N-k} + \frac{2}{N+1} \nonumber\\&- \frac{11}{3}\frac{1}{N+2} + \frac{1}{(N+2)^2}\Bigg\},
\end{align}
\begin{align}
\label{eq:LCD}
    \gamma_{N,k}^{\mathcal{D}, (1)}\biggr\vert_{\text{LC}} =& 4\:C_F^2\Bigg\{\frac{\Big(S_{1}(N)-S_{1}(k)\Big)^2}{N-k}\nonumber\\&+\Big(S_{1}(N)-S_{1}(k)\Big)\Big(S_{1}(N-k)-S_{1}(k)\Big)\Big(\frac{1}{N+2} -\frac{2}{N-k}\Big)\nonumber\\&+\frac{1}{2}\Big(S_{1}(N)-S_{1}(k)\Big)\Big(\frac{13}{3}\frac{1}{N-k}+\frac{4}{(N-k)^2}-\frac{2}{N+1}\frac{1}{k+1}  +\frac{2}{k+1}\frac{1}{N-k}\nonumber\\&-\frac{13}{3}\frac{1}{N+2}-\frac{2}{(N+2)^2}\Big)+\Big(S_{1}(N-k)-S_{1}(k)\Big)\Big(\frac{1}{N+1}  +\frac{1}{N+1}\frac{1}{k+1}\nonumber\\&-\frac{1}{k+1}\frac{1}{N+2}-\frac{1}{N+2}+\frac{1}{(N+2)^2}\Big) -\Big(S_{2}(N)-S_{2}(k)\Big)\Big(\frac{1}{N+2}-\frac{1}{N-k}\Big)\nonumber\\&-2S_{2}(k)\Big(\frac{1}{N+2}-\frac{1}{N-k}\Big) -\frac{67}{9}\frac{1}{N-k}  -\frac{53}{6}\frac{1}{N+1}-\frac{1}{(N+1)^2}\nonumber\\&-\frac{1}{(N+1)^2}\frac{1}{k+1}-\frac{1}{2}\frac{1}{N+1}\frac{1}{k+1}  +\frac{1}{2}\frac{1}{k+1}\frac{1}{N+2}  +\frac{293}{18}\frac{1}{N+2}-\frac{5}{3}\frac{1}{(N+2)^2}\nonumber\\&-\frac{1}{(N+2)^3}-\frac{1}{(N+2)^2}S_{1}(k) \Bigg\}
    \, .
\end{align}
The corresponding expression for the two-loop subleading-color anomalous dimensions in the derivative basis is currently not known. Nevertheless, using our computation of the operator matrix elements in combination with the consistency relation in Eq.(\ref{mainConj}), we derived closed-form expressions for the next-to-diagonal
\begin{align}
\label{eq:SLCntd}
    \gamma_{N,N-1}^{\D,(1)}\biggr\vert_{\text{SLC}} &= C_F\Bigg(C_F-\frac{C_A}{2}\Bigg)\Bigg(\frac{268}{9}-\frac{34}{3}\frac{1}{N}+\frac{4(-1)^N}{N}+\frac{4}{N^2}-\frac{4(-1)^N}{N^2}+\frac{160}{3}\frac{1}{N+1}\nonumber\\&-\frac{8(-1)^N}{N+1}-\frac{778}{9}\frac{1}{N+2}+\frac{4(-1)^N}{N+2}+\frac{32}{3}\frac{1}{(N+2)^2}-\frac{4(-1)^N}{(N+2)^2}-\frac{8}{(N+2)^3}\nonumber\\&-\frac{8(-1)^N}{(N+2)^3}+16S_{-2}(N)-\frac{16S_{-2}(N)}{N+2}\Bigg)
\end{align}
and for the last column
\begin{align}
\label{eq:SLClc}
    \gamma_{N,0}^{\D,(1)}\biggr\vert_{\text{SLC}} &=C_F\Big(C_F-\frac{C_A}{2}\Big)\Bigg(\frac1{N+1}\Bigg[ \frac{136}{3} + 16S_{{}1}(N) \Bigg]+  \frac{1}{(N+2)^2} \Bigg[ \frac{{}20}{{}3} + 8S_{{}1}(N) \Bigg] \nonumber\\&+  \frac1{N+2} \Bigg[  - \frac{676}{9} + \frac{{}20}{{}3}S_{{}1}(N) - 8
         S_{{}2}(N) \Bigg]-\frac{8}{N^2}  S_{{}1}(N) \nonumber\\&+  \frac1{N} \Bigg[ \frac{268}{9} - \frac{{}68}{{}3}S_{{}1}(N) + 8
         S_{{}2}(N) \Bigg]\Bigg).
\end{align}
This provides enough information to fix the spin-four ADM
\begin{equation}
\label{eq:SLCN4}
    \hat{\gamma}^{\D,(1)}(N=4)\biggr\vert_{\text{SLC}} = C_F\Big(C_F-\frac{C_A}{2}\Big)\begin{pmatrix}
   -\frac{32314}{675} & \frac{2782}{225} & \frac{211}{50} & \frac{1916}{675} \\[6pt]
0 & -\frac{1070}{27} & \frac{106}{9} & \frac{97}{18} \\[6pt]
0 & 0 & -\frac{752}{27} & \frac{376}{27} \\[6pt]
0 & 0 & 0 & 0
    \end{pmatrix}.
\end{equation}
The spin-two and -three results agree with previous computations~\cite{Gracey:2009da}. However, more information is needed to go beyond $N=4$. Moreover, the appearance of negative-index harmonic sums makes the sums in the consistency relation in Eq.(\ref{mainConj}) more difficult to evaluate analytically as compared to the leading-color case. Hence the fully analytic treatment of the subleading-color anomalous dimensions is left for a future publication. In the current work we show how to generate the subleading-color ADMs for fixed $N$ values based on the expressions in the Gegenbauer basis. For this, consider the general expression of the Gegenbauer anomalous dimensions at two loops, cf. Eq.(\ref{eq:geg2L}),
\begin{equation}
    \gamma_{N,k}^{\mathcal{G}, (1)} \,=\, \delta_{N,k}\gamma_{k,k}^{(1)}
     -\frac{\gamma_{N,N}^{(0)}-\gamma_{k,k}^{(0)}}{a(N,k)}\Bigg\{-2(2k+3)\Big(\beta_0+\gamma_{k,k}^{(0)}\Big)\vartheta_{N,k}+2\Delta_{N,k}^{\G,(0)}\Bigg\}
    \, .
\end{equation}

\begin{figure}
           \centering
           \includegraphics[width=0.47\textwidth]{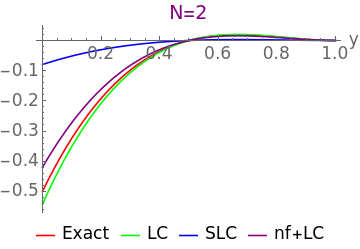}
           \includegraphics[width=0.47\textwidth]{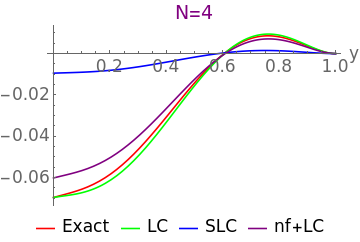}
           \includegraphics[width=0.47\textwidth]{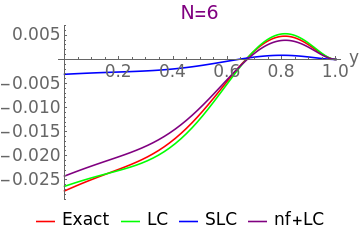}
           \includegraphics[width=0.47\textwidth]{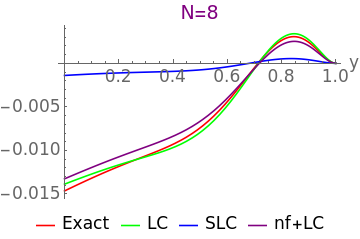}
           \includegraphics[width=0.47\textwidth]{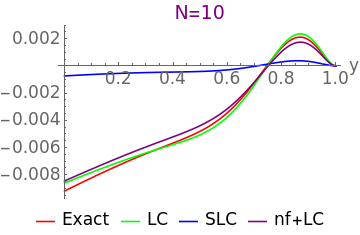}
           \caption{Comparison of the exact two-loop Wilson anomalous dimensions with the leading- and subleading-color approximations. As expected, the leading-color limit provides a good approximation to the full result. However, adding the leading-$n_f$ expression actually worsens the approximation a bit.}
           \label{fig:fig1}
\end{figure}

Let us now focus on the subleading-color part of the two-loop ADM. We see from the expression above that this is completely determined by the $\beta_0$ contribution\footnote{The subleading-color part corresponds to $-2$ times the terms proportional to $C_F C_A$.}
\begin{equation}
    \gamma_{N,k}^{\mathcal{G}, (1)}\biggr\vert_{\text{SLC}} \,=\, \delta_{N,k}\gamma_{k,k}^{(1)}\biggr\vert_{\text{SLC}} +\frac{2(\gamma_{N,N}^{(0)}-\gamma_{k,k}^{(0)})}{a(N,k)}[-2(2k+3)\beta_0\big\vert_{C_A}\vartheta_{N,k}]
\end{equation}
where we only need to take into account the term in $\beta_0$ proportional to $C_A$. It is clear then that the subleading-color part of the Gegenbauer ADM has the same functional form as the leading-$n_f$ part, just with different numeric and color factors. The leading-$n_f$ approximation of the non-diagonal elements was computed in~\cite{Moch:2021cdq} to be
\begin{align}
    \gamma_{N,k}^{\mathcal{G}, (1)}\biggr\vert_{n_f} = \frac{8}{3}\frac{n_f C_F}{a(N,k)}\vartheta_{N,k}&\Bigg\{-2\Big(S_{1}(N)-S_{1}(k)\Big)(2k+3)-(2k+3)\Big(\frac{1}{N+1}+\frac{1}{N+2}\Big)\nonumber\\&+4+\frac{1}{k+1}-\frac{1}{k+2} \Bigg\}
\end{align}
and hence the expression for the subleading-color part becomes
\begin{align}
    \gamma_{N,k}^{\mathcal{G}, (1)}\biggr\vert_{\text{SLC}} = \frac{88}{3}C_F\Big(C_F-\frac{C_A}{2}\Big)\frac{\vartheta_{N,k}}{a(N,k)}&\Bigg\{-2\Big(S_{1}(N)-S_{1}(k)\Big)(2k+3)-(2k+3)\Big(\frac{1}{N+1}\nonumber\\&+\frac{1}{N+2}\Big)+4+\frac{1}{k+1}-\frac{1}{k+2} \Bigg\}.
\end{align}
The corresponding anomalous dimensions in the derivative basis are then determined by the basis transformation as
\begin{equation}
\label{eq:SLC}
  \gamma_{N,k}^{\D,(1)}\biggr\vert_{\text{SLC}}  = \frac{(-1)^k(N+1)!}{(k+1)!}\sum_{l=k}^{N}(-1)^l\binom{N}{l}\frac{l!(3+2l)}{(N+l+3)!}\sum_{j=k}^{l}\binom{j}{k}\frac{(j+k+2)!}{j!}\gamma_{l,j}^{\G,(1)}\biggr\vert_{\text{SLC}}.
\end{equation}
For $k=N-1$ and $k=0$ we find full agreement with Eqs.(\ref{eq:SLCntd}) and (\ref{eq:SLClc}). Furthermore, using Eq.(\ref{eq:ERBL2}), we find full agreement with the $x$-space expression presented in~\cite{Mikhailov:1984ii}. To illustrate our results, we compare the leading-color anomalous dimensions to the now available full expressions. Specifically we plot the following quantities
\begin{align}
    &\frac{(1-y)^2}{\gamma_{N,N}^{(1)}}\sum_{k=0}^{N}\gamma_{N,k}^{\D,(1)}y^k,\\&
    \frac{(1-y)^2}{\gamma_{N,N}^{(1)}}\sum_{k=0}^{N}\gamma_{N,k}^{\D,(1)}\biggr\vert_{\text{LC}}y^k,\\&
     \frac{(1-y)^2}{\gamma_{N,N}^{(1)}}\sum_{k=0}^{N}\Bigg(\gamma_{N,k}^{\D,(1)}\biggr\vert_{\text{LC}}+\gamma_{N,k}^{\D,(1)}\biggr\vert_{n_f}\Bigg)y^k,\\&
    \frac{(1-y)^2}{\gamma_{N,N}^{(1)}}\sum_{k=0}^{N}\gamma_{N,k}^{\D,(1)}\biggr\vert_{\text{SLC}}y^k.
\end{align}
The prefactor is introduced for normalization purposes. The sum however is phenomenologically interesting as it appears in the evolution equations of non-forward parton distributions and hadronic wavefunctions, cf.~Eqs.(\ref{eq:ERBL1}) and (\ref{eq:ERBL2}).
We plot the results in QCD ($C_A=3$ and $C_F=4/3$) and set $n_f=3$, see Fig.\ref{fig:fig1}. Plots are shown for spin-$N$ operators with $N={2,4,6,8,10}$. The corresponding expressions for the anomalous dimensions, in $SU(n_c)$, can be found in Appendix \ref{ap:L2ADM}. As expected, the exact anomalous dimensions are approximated well by the leading-color limit. Note however that including the leading-$n_f$ expression actually leads to a slightly worse approximation of the exact result\footnote{The quality of the approximation improves however for higher values of $n_f$.}.\newline

The analysis presented above can be repeated to compute the two-loop subleading-color part of the transversity anomalous dimensions. The corresponding leading-$n_f$ expression in the Gegenbauer basis was computed in~\cite{VanThurenhout:2022nmx} to be
\begin{align}
    \begin{split}
        \gamma_{N,k}^{T,\mathcal{G},(1)}\biggr\vert_{n_f} = -\frac{16}{3}\frac{n_fC_F}{a(N,k)}\vartheta_{N,k}\Bigg\{(3+2k)\Bigg(S_1(N)-S_1(k)+\frac{1}{N+1}\Bigg)-\frac{1}{k+1}-2\Bigg\}.
    \end{split}
\end{align}
Hence the expression for the subleading-color part becomes
\begin{align}
    \begin{split}
        \gamma_{N,k}^{T,\mathcal{G},(1)}\biggr\vert_{\text{SLC}} = -\frac{176}{3}\frac{\vartheta_{N,k}}{a(N,k)}C_F\Big(C_F-\frac{C_A}{2}\Big)\Bigg\{(3+2k)\Bigg(S_1(N)-S_1(k)+\frac{1}{N+1}\Bigg)-\frac{1}{k+1}-2\Bigg\}
    \end{split}
\end{align}
such that in the derivative basis we have
\begin{equation}
\label{eq:SLCTr}
  \gamma_{N,k}^{T,\D,(1)}\biggr\vert_{\text{SLC}}  = \frac{(-1)^k(N+1)!}{(k+1)!}\sum_{l=k}^{N}(-1)^l\binom{N}{l}\frac{l!(3+2l)}{(N+l+3)!}\sum_{j=k}^{l}\binom{j}{k}\frac{(j+k+2)!}{j!}\gamma_{l,j}^{T,\G,(1)}\biggr\vert_{\text{SLC}}.
\end{equation}
To showcase some explicit results, we repeat the analysis performed above for the Wilson operators. Specifically, for the spin-$N$ transversity operators, with $N=2,4,6,8,10$, we plot
\begin{align}
    &\frac{(1-y)^2}{\gamma_{N,N}^{T,(1)}}\sum_{k=0}^{N}\gamma_{N,k}^{T,\D,(1)}y^k,\\&
    \frac{(1-y)^2}{\gamma_{N,N}^{T,(1)}}\sum_{k=0}^{N}\gamma_{N,k}^{T,\D,(1)}\biggr\vert_{\text{LC}}y^k,\\&
    \frac{(1-y)^2}{\gamma_{N,N}^{(1)}}\sum_{k=0}^{N}\Bigg(\gamma_{N,k}^{T,\D,(1)}\biggr\vert_{\text{LC}}+\gamma_{N,k}^{T,\D,(1)}\biggr\vert_{n_f}\Bigg)y^k,\\&
    \frac{(1-y)^2}{\gamma_{N,N}^{T,(1)}}\sum_{k=0}^{N}\gamma_{N,k}^{T,\D,(1)}\biggr\vert_{\text{SLC}}y^k
\end{align}
comparing the \textit{full}\footnote{We use here that the one-loop conformal anomaly is the same for the Wilson and transversity anomalous dimensions.} two-loop anomalous dimensions with the leading- and subleading-color approximations, see Fig.\ref{fig:fig2}. The corresponding expressions in $SU(n_c)$ can be found in Appendix \ref{ap:L2ADMTr}. Again we see that the leading-color limit provides a good approximation to the full result (although the agreement is not as good as in the Wilson case). Furthermore, adding the leading-$n_f$ expression actually worsens the approximation a bit\footnote{But as before, the quality of the approximation improves for higher values of $n_f$.}.
\begin{figure}
           \centering
           \includegraphics[width=0.47\textwidth]{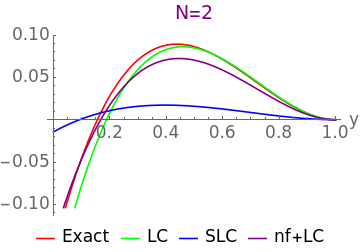}
           \includegraphics[width=0.47\textwidth]{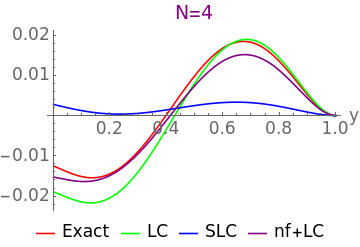}
           \includegraphics[width=0.47\textwidth]{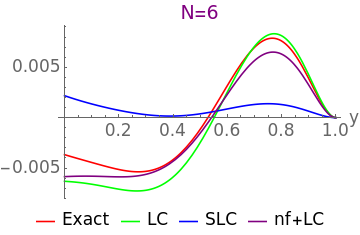}
           \includegraphics[width=0.47\textwidth]{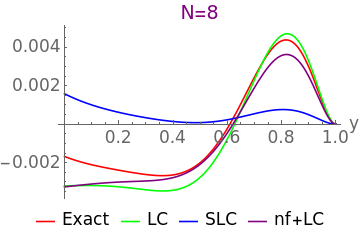}
           \includegraphics[width=0.47\textwidth]{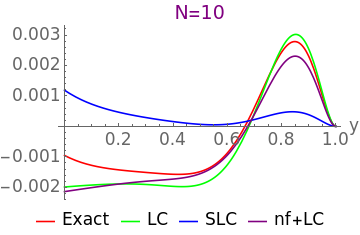}
           \caption{Comparison of the exact two-loop transversity anomalous dimensions with the leading- and subleading-color approximations. As expected, the leading-color limit provides a good approximation to the full result (although the agreement is not as good as in the Wilson case). However, adding the leading-$n_f$ expression actually worsens the approximation a bit.}
           \label{fig:fig2}
\end{figure}

\section{Validating conformal anomaly computations}
\label{sec:anomaly}
Within conformal schemes the conformal anomaly is one of the main ingredients (and bottlenecks) in the determination of the anomalous dimension matrices of composite operators. At leading order it vanishes, reflecting the fact that conformal symmetry is exact at this level. As discussed above, this implies a diagonal ADM. However, at higher orders conformal symmetry is lost, i.e. the conformal anomaly becomes non-zero and the ADM develops non-zero off-diagonal elements. At the one-loop level, the conformal anomaly is known to be of the form\footnote{There is a factor of two difference here with respect to the expression in \cite{Braun:2017cih} due to different conventions in the definition of the anomalous dimensions.}~\cite{Braun:2017cih,Braun:2016qlg,Braun:2014vba}
\begin{equation}
\label{eq:anomaly}
    \Delta_{N,k}^{\G,(0)} = 2C_F(2k+3)a(N,k)\Bigg(\frac{A_{N,k}-S_1(N+1)}{(k+1)(k+2)}+\frac{2A_{N,k}}{a(N,k)}\Bigg)\vartheta_{N,k}
\end{equation}
with
\begin{equation}
    A_{N,k} = S_{1}\Bigg(\frac{N+k+2}{2}\Bigg)-S_{1}\Bigg(\frac{N-k-2}{2}\Bigg)+2S_{1}(N-k-1)-S_{1}(N+1).
\end{equation}
The two-loop conformal anomaly is known, however no closed-form expression exists thus far~\cite{Braun:2017cih}. We now present a method to independently check existing computations of the conformal anomaly, based on our similarity transformation. In what follows we will focus on the one-loop anomaly, but the generalization to higher orders is straightforward. Consider the expression for the off-diagonal part of the two-loop Gegenbauer ADM, cf.~Eq.(\ref{eq:geg2L})
\begin{equation}
    \gamma_{N,k}^{\mathcal{G}, (1)} \,=\, -\frac{\gamma_{N,N}^{(0)}-\gamma_{k,k}^{(0)}}{a(N,k)}\Bigg\{-2(2k+3)\Big(\beta_0+\gamma_{k,k}^{(0)}\Big)\vartheta_{N,k}+2\Delta_{N,k}^{\G,(0)}\Bigg\}
    \, .
\end{equation}
Let us now focus on the leading-color limit of this expression, which can be written as
\begin{equation}
\label{eq:tslv}
     \gamma_{N,k}^{\mathcal{G}, (1)}\biggr\vert_{\text{LC}} \,=\, \frac{\gamma_{N,N}^{(0)}-\gamma_{k,k}^{(0)}}{a(N,k)}\Bigg\{(2k+3)\Big(2\gamma_{k,k}^{(0)}+\frac{44}{3}C_F\Big)\vartheta_{N,k}-2\Delta_{N,k}^{\G,(0)}\Bigg\}
    \, .
\end{equation}
Solving Eq.(\ref{eq:tslv}) to the one-loop anomaly leads to
\begin{equation}
    \Delta_{N,k}^{\G,(0)} = (2k+3)\Bigg(\gamma_{k,k}^{(0)}+\frac{22}{3}C_F\Bigg)-\frac{(N-k)(N+k+3)}{2(\gamma_{N,N}^{(0)}-\gamma_{k,k}^{(0)})}\gamma_{N,k}^{\G,(1)}\biggr\vert_{\text{LC}}
\end{equation}
in which we substituted the expression for $a(N,k)$, cf.~Eq.(\ref{eq:defa}). Now we can use our basis transformation, Eq.(\ref{eq:basisTransIn}), to write the right-hand-side in terms of the leading-color anomalous dimensions in the derivative basis
\begin{equation}
\label{eq:DelG}
    \Delta_{N,k}^{\G,(0)} = (2k+3)\Bigg(\gamma_{k,k}^{(0)}+\frac{22}{3}C_F\Bigg)-\frac{(N-k)(N+k+3)}{2(\gamma_{N,N}^{(0)}-\gamma_{k,k}^{(0)})}R_{N,l}\Tilde{R}_{j,k}\gamma_{l,j}^{\D,(1)}\biggr\vert_{\text{LC}}
\end{equation}
with $R_{N,k}$ and $\Tilde{R}_{N,k}$ given by Eqs.(\ref{eq:R}) and (\ref{eq:RTilde}) respectively. The sum over $l$ and $j$ is understood, cf.~Eq.(\ref{eq:DtoG}). The two-loop leading-color anomalous dimensions in the derivative basis were computed in~\cite{Moch:2021cdq}, cf.~Eq.(\ref{eq:LCD}). At present it is unclear if the sums appearing in Eq.(\ref{eq:DelG}) can be performed analytically. Nevertheless, using e.g. the \textit{S{\small UMMER}}~\cite{Vermaseren:1998uu} package in \textit{F{\small ORM}}~\cite{Vermaseren:2000nd,Kuipers:2012rf}, they can be efficiently evaluated for fixed $(N,k)$ values. The result agrees with the expression in~\cite{Braun:2017cih}, cf.~Eq.(\ref{eq:anomaly}). Note that, in Eq.(\ref{eq:DelG}), the one-loop anomaly is written in terms of two-loop anomalous dimensions. More generally, the $L$-loop conformal anomaly in the Gegenbauer basis will be determined by the $(L+1)$-loop leading-color anomalous dimensions in the derivative basis. While this may not seem ideal, we believe that Eq.(\ref{eq:DelG}) and its higher-order generalizations will still be useful for two reasons.
\begin{enumerate}
    \item The anomaly will be written in terms of sums of functions involving denominators and harmonic sums which, at least for fixed $(N,k)$ values, can be computed quickly and efficiently using modern computer algebra packages like \textit{S{\small UMMER}} in \textit{F{\small ORM}} or \textit{S{\small IGMA}} in \textit{M{\small ATHEMATICA}}.
    \item Although the leading-color anomalous dimensions in the derivative basis are one order higher than the anomaly itself, they should, at least in principle, be relatively simple to compute. The information necessary is comprised of the forward anomalous dimension (which is known completely to four loops in the leading-color approximation~\cite{Gross:1973ju, Floratos:1977au, Moch:2004pa,Moch:2017uml}) and the operator matrix elements, which serve as a boundary condition to our consistency relation, cf.~Eq.(\ref{mainConj}). These matrix elements can be computed using standard computer algebra techniques.
\end{enumerate}
Finally, to finish off this section, we use our basis transformation formula to compute the one-loop conformal anomaly in the derivative basis. We find
\begin{align}
    \Delta_{N,k}^{\D,(0)} &= \frac{(-1)^k(N+1)!}{(k+1)!}\sum_{l=k}^{N}(-1)^l\binom{N}{l}\frac{l!(3+2l)}{(N+l+3)!}\sum_{j=k}^{l}\binom{j}{k}\frac{(j+k+2)!}{j!}\Delta_{l,j}^{\G} \\ &= C_F\Bigg(\frac{k+1}{N-k-1}-\frac{1}{N-k}-S_1(N)+S_1(N-k)\Bigg)
\end{align}
if $k\neq N-1$ and $\Delta_{N,N-1}^{\D,(0)} = 0$. Of note is that this expression obeys the consistency relation Eq.(\ref{mainConj}), i.e. we have
\begin{equation}
    \Delta_{N,k}^{\D,(0)} = \sum_{j=k}^{N}(-1)^{k}\binom{j}{k}\sum_{l=j+1}^{N}(-1)^{l}\binom{l}{j}\Delta_{l,j}^{\D,(0)}.
\end{equation}

\section{One-loop pure gluon ADM}
\label{sec:gluon}
As an application of the gluonic similarity transformation, cf.~Eqs.(\ref{eq:gluon1}), (\ref{eq:gluon2}) and (\ref{eq:gluon3}), we compute the one-loop ADM $\hat{\gamma}_{g,g}^{\D}$ in the derivative basis. Generically, the mixing of quark and gluon operators in the flavor-singlet sector is written as a matrix equation of the form\footnote{We focus in this work exclusively on the gauge invariant operators, ignoring the additional mixing with non-gauge invariant and equation of motion operators.}
\begin{equation}
\label{eq:singlet}
    \begin{pmatrix}
        \Op^{q,\D} \\ \Op^{g,\D}
    \end{pmatrix} = 
    \begin{pmatrix}
        Z^{qq,\D} & Z^{qg,\D} \\
        Z^{gq,\D} & Z^{\:gg,\D}
    \end{pmatrix}
    \begin{pmatrix}
        [\Op^{q,\D}] \\ [\Op^{g,\D}]
        \end{pmatrix}.
\end{equation}
In the forward limit, the renormalization factors $Z^{ij,\D}$ in Eq.(\ref{eq:singlet}) are regular functions of the dimensional regulator $\eps$, the renormalization group parameters and the strong coupling $a_s$. However in the case of non-forward kinematics, for which one has to take into account mixing with the total-derivative operators in Eqs.(\ref{eq:singQ}) and (\ref{eq:singG}), the renormalization factors, and hence the anomalous dimensions, become matrices themselves. In the present work we are interested in the $\hat{Z}^{\:gg,\D}$ renormalization factor, or more specifically in the associated ADM at the one-loop level
\begin{equation}
    \hat{Z}^{\:gg,\D} = \mathbb{1} + \frac{a_s}{\eps}\hat{\gamma}^{\:gg,\D,(0)}+O(a_s^2).
\end{equation}
As was the case for the quark flavor-non-singlet operators, the diagonal elements of the gluon ADM are simply the forward anomalous dimensions~\cite{Balitsky:1987bk,Braunschweig:1987dr}
\begin{equation}
    \gamma_{N,N}^{\:gg,(0)} = C_A\Bigg(4S_{1}(N+1)+\frac{4}{N+3}-\frac{4}{N+2}+\frac{4}{N+1}-\frac{4}{N}-\frac{11}{3}\Bigg)+\frac{2}{3}n_f.
\end{equation}
These do not depend on the basis chosen for the total-derivative operators. Note the pole at $N=0$ in this expression. Using now our gluonic basis transformation and the fact that, in the Gegenbauer basis, the gluon ADM is diagonal, we obtain the following representation for the off-diagonal elements of the ADM in the derivative basis
\begin{equation}
    \gamma_{N,k}^{\:gg,(0),\D} = \frac{(-1)^{k}(N+1)!}{N(k+1)!}\sum_{l=k}^{N}(-1)^l\binom{N}{l}\binom{l-1}{k-1}\frac{(3+2l)l!(l+k+2)!}{(N+l+3)!(l-1)!}\gamma_{l,l}^{\:gg,(0)}.
\end{equation}
Evaluating the sum we find
\begin{equation}
\label{eq:Ggg}
    \gamma_{N,k}^{\:gg,(0),\D} = C_A\Bigg(\frac{-4(2+k)}{N+3}+\frac{4(3+2k)}{N+2}-\frac{6k}{N+1}+\frac{2(k-1)}{N}-\frac{2}{N-k}\Bigg).
\end{equation}
Note the appearance of $k$-factors in the numerators, which were absent in the flavor-non-singlet sector. For illustration, we explicitly present the spin-five ADM
\begin{align}
    \hat{\gamma}_{N=5}^{\:gg,\D,(0)} &= C_A\begin{pmatrix}
        \gamma_{4,4}^{\:gg,(0)} & \gamma_{4,3}^{\:gg,(0),\D} &\gamma_{4,2}^{\:gg,(0),\D} & \gamma_{4,1}^{\:gg,(0),\D} \\[6pt]
        0 & \gamma_{3,3}^{\:gg,(0)} & \gamma_{3,2}^{\:gg,(0),\D} & \gamma_{3,1}^{\:gg,(0),\D} \\[6pt]
        0 & 0 &\gamma_{2,2}^{\:gg,(0)} & \gamma_{2,1}^{\:gg,(0),\D} \\[6pt]
        0 & 0 & 0 & \gamma_{1,1}^{\:gg,(0)}
    \end{pmatrix}\\
    \nonumber\\&= C_A\begin{pmatrix}
        \frac{181}{35} & -\frac{51}{35} & -\frac{109}{210} & -\frac{26}{105} \\[6pt]
        0 & \frac{21}{5} & -\frac{7}{5} & -\frac{1}{2} \\[6pt]
        0 & 0 & \frac{14}{5} & -\frac{7}{5} \\[6pt]
        0 & 0 & 0 & 0
    \end{pmatrix}
    +\frac{2}{3}n_f \mathbb{1}_{4\times 4}.
\end{align}
Note that the spin-$N$ ADM is now an $(N-1)\times(N-1)$ matrix because of the absence of a spin-one operator. Also, the gluonic part of the spin-two anomalous dimension vanishes as the corresponding operator is the gluonic part of the QCD energy-momentum tensor, which is conserved.\newline

Our result in Eq.(\ref{eq:Ggg}) can be checked by comparing it to a computation performed in another basis for the operators, which is due to B. Geyer and friends~\cite{Geyer:1982fk,Blumlein:1999sc}. This type of comparison was used before to check the expression of the one-loop ADM for the flavor-non-singlet quark operators, see~\cite{Moch:2021cdq}. The gluon operators in this basis are of the form
\begin{equation}
\label{eq:geyer}
    \Op_{N,k}^{\text{Geyer}} = F_{\mu\Delta}(\cev D \cdot \Delta+\Delta \cdot\Vec{D})^{N-k}(\cev D \cdot \Delta-\Delta \cdot\Vec{D})^{k-1}F^{\mu\Delta}.
\end{equation}
In what follows we focus on the diagonal operator, $\Op_{N,N}^{\text{Geyer}}$, which obeys the following evolution equation
\begin{equation}
    \mu^2 \frac{\text{d}}{\text{d}\mu^2}\mathcal{O}_{N+1,N+1}^{\text{Geyer}} = \gamma^{\:gg,(0)}_{N+1,N+1}+\sum_{k=0}^{N}\frac{1-(-1)^k}{2}\gamma_{N+1,k}^{(0),\text{Geyer}}[\mathcal{O}^{\text{Geyer}}_{k,N+1}].
\end{equation}
We now follow the same steps as those presented in~\cite{Moch:2021cdq} for the quark operators. The operators in Eq.(\ref{eq:geyer}) are first written in terms of those in the derivative basis. Then one introduces the renormalization and sets the renormalized operators to one, which is valid at the one-loop level. The result is a relation between the anomalous dimensions in the Geyer basis and the last column ($k=1$) in the derivative one, which reads
\begin{equation}
    \gamma^{\:gg,(0)}_{N+1,N+1}+\sum_{k=0}^{N}\frac{1-(-1)^k}{2}\gamma_{N+1,k}^{(0),\text{Geyer}} = (-1)^N \sum_{l=0}^{N}(-1)^{l}\:2^{l}\binom{N}{l}\gamma_{l+1,1}^{\:gg,\D,(0)}
\end{equation}
and is valid for even $N$. We have checked that our result in Eq.(\ref{eq:Ggg}), which gives
\begin{equation}
    \gamma_{N,1}^{\:gg,\D,(0)} = C_A\Bigg(-\frac{12}{N+3}+\frac{20}{N+2}-\frac{6}{N+1}-\frac{2}{N-1}\Bigg)
\end{equation} 
for the last column, obeys this relation.

\section{Conclusion and outlook}
\label{sec:conclusion}
We have derived the explicit form of the similarity transformation between two operator bases which are often used in the renormalization of leading-twist operators relevant for the description of exclusive scattering processes. The first basis writes the operators in terms of Gegenbauer polynomials, while the second is based on counting powers of derivatives. Two important consequences of this transformation are (a) the non-forward anomalous dimensions in the derivative basis admit, at the one-loop level, a representation in terms of the forward anomalous dimensions and (b) a direct physical interpretation of a consistency relation between the anomalous dimensions in the derivative basis, previously derived from the renormalization structure of the operators in the chiral limit, in terms of CP-symmetry. An extension of this procedure to higher-twist operators would be interesting but is left for future studies.\newline

We have also presented several applications of our basis transformation. As a first application, we extended a previous calculation of the all-order anomalous dimensions, in the limit of a large number of quark flavors $n_f$, in the derivative basis to corresponding expressions in the Gegenbauer basis. In the article we focused on the spin-five operators, which were expanded to sixth order in the strong coupling. However, generalization to higher-spin operators and higher orders is straightforward. Next, we presented the anomalous dimensions for operators of spin $N=2,4,6,8,10$ to two-loop accuracy in full QCD in the derivative basis. For $N=2$ we find full agreement with the previously computed expression. The results for $N>2$, which previously were only known in the large-$n_f$ and large-$n_c$ approximations, are new and provide a check of the expectation that the large-$n_c$ approximation is actually a good measure for the full anomalous dimensions, which is useful for phenomenology. This analysis was performed for both the Wilson and transversity operators. Additionally, after performing a scheme transformation to the RI scheme, these results will also be useful
in non-perturbative lattice studies of the hadron structure. The derivation of a closed-form expression for the subleading-color part of the anomalous dimensions is in progress. The latter would be useful for the computation of the three-loop anomalous dimensions. The third application of our new basis transformation formula is a novel way to validate computations of the conformal anomaly, which is an important ingredient for the determination of the anomalous dimensions in conformal schemes. We show that the conformal anomaly at $L$-loop accuracy can be computed from the leading-color anomalous dimensions, in the derivative basis, at the ($L+1$)-loop level. While we do need to go one order in $\alpha_s$ higher, we nevertheless believe the method to be useful, as the leading-color anomalous dimensions in the derivative basis are relatively straightforward to compute using the consistency relation between the anomalous dimensions in combination with computer algebra techniques. Another advantage is that our method leads to a representation of the conformal anomaly in terms of sums. The latter can be evaluated efficiently using currently available computer algebra packages, like \textit{S{\small UMMER}} in \textit{F{\small ORM}} or \textit{S{\small IGMA}} in \textit{M{\small ATHEMATICA}}. We have applied our method to provide an independent check of the one-loop anomaly, based on the known two-loop anomalous dimensions in the large-$n_c$ limit. The computation of the three-loop large-$n_c$ anomalous dimensions in the derivative basis, and the corresponding check of the two-loop conformal anomaly, is left for a future study.\newline

Finally, we have derived the similarity transformation for the leading-twist gluon operators, which is important for the renormalization of composite operators in the flavor-singlet sector. As an application we computed the one-loop pure gluonic ($\gamma^{\:gg}$) anomalous dimensions in non-forward kinematics. It would be interesting to extend these results to both the non-diagonal elements of the singlet mixing matrix ($\gamma^{\:qg}$, $\gamma^{\:gq}$) and to higher orders in perturbation theory. We leave these considerations for future studies. 

\subsection*{Acknowledgements}
The author would like to thank A. Manashov and G. Somogyi for useful discussions and for valuable comments on the manuscript. This work has been supported by grant K143451 of the National Research, Development and Innovation Fund in Hungary.

\bigskip

\appendix

 ---------------------------------------------------------------------

\renewcommand{\theequation}{\ref{sec:appA}.\arabic{equation}}
\setcounter{equation}{0}
\renewcommand{\thefigure}{\ref{sec:appA}.\arabic{figure}}
\setcounter{figure}{0}
\renewcommand{\thetable}{\ref{sec:appA}.\arabic{table}}
\setcounter{table}{0}
\section{All-order spin-five anomalous dimensions in the large-$n_f$ limit}
\label{sec:appA}
In this appendix, we collect the all-order expressions for the elements of the spin-five ADM. We show the results both in the derivative and in the Gegenbauer basis, where the latter are derived from the first using the basis transformation in Eq.(\ref{eq:basisTransIn}). Furthermore, in both bases we present the anomalous dimensions for both the Wilson and the transversity operators.

\subsection{Results in the derivative basis}
In the derivative basis, the spin-five ADM has the following form
\begin{equation}
    \begin{pmatrix}
        \gamma_{4,4} & \gamma_{4,3}^{\D} & \gamma_{4,2}^{\D} & \gamma_{4,1}^{\D} & \gamma_{4,0}^{\D} \\[6pt]
        0 & \gamma_{3,3} & \gamma_{3,2}^{\D} & \gamma_{3,1}^{\D} & \gamma_{3,0}^{\D} \\[6pt]
        0 & 0 & \gamma_{2,2} & \gamma_{2,1}^{\D} & \gamma_{2,0}^{\D} \\[6pt]
        0 & 0 & 0 & \gamma_{1,1} & \gamma_{1,0}^{\D} \\[6pt]
        0 & 0 & 0 & 0 & \gamma_{0,0}
    \end{pmatrix}.
\end{equation}
For completeness, we also present the corresponding expressions for the forward anomalous dimensions, i.e. the diagonal elements of the ADM. These were computed before in~\cite{Gracey:1994nn} and~\cite{Gracey:2003mr} by J. Gracey. We repeat here that these diagonal elements are independent of the basis chosen for the total-derivative operators.

\subsubsection{Wilson operators}
To lighten the notation we define the following auxiliary functions
\begin{align}
    &\mathcal{F}_4(a_s,n_f) = \frac{2^{4
-\frac{4 a n_f}{3}
}}{\pi ^{3/2}}\frac{\Gamma \Big(
        \frac{5}{2}
        -\frac{2 a n_f}{3}
\Big) \sin \Big(
        \frac{2 a n_f \pi }{3}\Big)}{27 n_f \Gamma \
\Big(
        7
        -\frac{2 a n_f}{3}
\Big)}, \\&
    \mathcal{F}_3(a_s,n_f) = -\frac{2^{3
-\frac{4 a_s n_f}{3}
}}{\pi ^{3/2}}\frac{\Gamma \Big(
        \frac{5}{2}
        -\frac{2 a_s n_f}{3}
\Big) \sin \Big(
        \frac{2 a_s n_f \pi }{3}\Big)}{9 n_f \Gamma \Big(
        6
        -\frac{2 a_s n_f}{3}
\Big)} .
\end{align}
We then calculated the elements of the spin-five Wilson ADM to be
\begin{align}
\label{eq:wFirst}
    &\gamma_{4,4} = \frac{2}{3}(4 a_s n_f-21) (2 a_s n_f (324 + a_s n_f (4 a_s n_f-63))-1053)\mathcal{F}_4(a_s,n_f), \\&
    \gamma_{4,3}^{\D} = -\frac{2}{3}(a_s n_f-3) (2 a_s n_f-9) (225 + 8 a_s n_f (a_s n_f-9))\mathcal{F}_4(a_s,n_f), \\&
    \gamma_{4,2}^{\D} = 18 (a_s n_f-3) (30 + a_s n_f (2 a_s n_f-13))\mathcal{F}_4(a_s,n_f), \\&
    \gamma_{4,1}^{\D} = -6 (135 + 4 a_s n_f (4 a_s n_f-21))\mathcal{F}_4(a_s,n_f), \\&
    \gamma_{4,0}^{\D} = 135 (a_s n_f-3)\mathcal{F}_4(a_s,n_f), \\&
    \gamma_{3,3} = -1413 + 2 a_s n_f (483 + 4 a_s n_f (2 a_s n_f-27))\mathcal{F}_3(a_s,n_f), \\&
    \gamma_{3,2}^{\mathcal{D}} = -4(a_sn_f-3)[36+a_sn_f(2a_sn_f-15)]\mathcal{F}_3(a_s,n_f),\\&
    \gamma_{3,1}^{\mathcal{D}} = 9(18+a_sn_f(2a_sn_f-11))\mathcal{F}_3(a_s,n_f),\\&
    \gamma_{3,0}^{\mathcal{D}} = -24(a_sn_f-3)\mathcal{F}_3(a_s,n_f), \\&
    \gamma_{2,2} = \frac{1}{3} (15 - 4 a_s n_f)^2 (2 a_s n_f-15)\mathcal{F}_3(a_s,n_f), \\&
    \gamma_{2.1}^{\D} = \frac{1}{3} (15 - 2 a_s n_f) (81 + 8 a_s n_f (a_s n_f-6))\mathcal{F}_3(a_s,n_f), \\&
    \gamma_{2,0}^{\D} = 3 (a_s n_f-3) (2 a_s n_f-15)\mathcal{F}_3(a_s,n_f), \\&
    \gamma_{1,1} = \frac{8}{3} (a_s n_f-6) (a_s n_f-3) (2 a_s n_f-15)\mathcal{F}_3(a_s,n_f), \\&
    \label{eq:WLast}
    \gamma_{1,0}^{\D} = -\frac{4}{3} (a_s n_f-6) (a_s n_f-3) (2 a_s n_f-15)\mathcal{F}_3(a_s,n_f).
\end{align}
Note that, for the Wilson operators, $\gamma_{0,0}=0$ to all orders in perturbation theory as it corresponds to the anomalous dimension of a conserved current. 

\subsubsection{Transversity operators}
Defining
\begin{align}
   &\mathcal{F}^T_4(a_s,n_f) = \frac{2^{4
-\frac{4 a_s n_f}{3}
}}{\pi ^{3/2}}\frac{\Gamma \Big(
        \frac{5}{2}
        -\frac{2 a_s n_f}{3}
\Big) \sin \Big(
        \frac{2 a_s n_f \pi }{3}\Big)}{81 n_f (-3
+2 a_s n_f
) \Gamma \Big(
        7
        -\frac{2 a_s n_f}{3}
\Big)} , \\&
\mathcal{F}_{3}^{T}(a_s,n_f) = -\frac{2^{3
-\frac{4 a_s n_f}{3}
}}{\pi ^{3/2}}\frac{\Gamma \Big(
        \frac{5}{2}
        -\frac{2 a_s n_f}{3}
\Big) \sin \Big(
        \frac{2 a_s n_f \pi }{3}\Big)}{27 n_f (-3
+2 a_s n_f
) \Gamma \Big(
        6
        -\frac{2 a_s n_f}{3}
\Big)} , \\&
\mathcal{F}^T_2(a_s,n_f)=-\frac{2^{3
-\frac{4 a_s n_f}{3}
}}{\pi ^{3/2}}\frac{\Gamma \Big(
        \frac{5}{2}
        -\frac{2 a_s n_f}{3}
\Big) \sin \Big(
        \frac{2 a_s n_f \pi }{3}\Big)}{9 n_f (-3
+2 a_s n_f
) \Gamma \Big(
        5
        -\frac{2 a_s n_f}{3}
\Big)}, \\&
\mathcal{F}^T_1(a_s,n_f)=\frac{2^{2
-\frac{4 a_s n_f}{3}
}}{\pi ^{3/2}}\frac{\Gamma \Big(
        \frac{5}{2}
        -\frac{2 a_s n_f}{3}
\Big) \sin \Big(
        \frac{2 a_s n_f \pi }{3}\Big)}{3 n_f (-3
+2 a_s n_f
) \Gamma \Big(
        4
        -\frac{2 a_s n_f}{3}
\Big)} 
\end{align}
we find
\begin{align}
\label{eq:TFirst}
    &\gamma_{4,4}^{T} = 9 (-3
+2 a_s n_f
) (69
+a_s n_f (-23
+2 a_s n_f
)
) (72
+a_s n_f (-35
+4 a_s n_f
))\mathcal{F}^T_4(a_s,n_f), \\&
    \gamma_{4,3}^{T,\D} = -8(-3
+a_s n_f
) (-9
+2 a_s n_f
) (-3
+2 a_s n_f
) (54
+a_s n_f (-15
+2 a_s n_f
)
)\mathcal{F}^T_4(a_s,n_f), \\&
    \gamma_{4,2}^{T,\D} = 54 (-3
+a_s n_f
) (-3
+2 a_s n_f
) (27
+a_s n_f (-9
+2 a_s n_f
))\mathcal{F}^T_4(a_s,n_f), \\&
    \gamma_{4,1}^{T,\D} = -72 (-3
+2 a_s n_f
) (27
+4 a_s n_f (-3
+a_s n_f
)
)\mathcal{F}^T_4(a_s,n_f), \\&
    \gamma_{4,0}^{T,\D} = 81 (27
+a_s n_f (-21
+10 a_s n_f
))\mathcal{F}^T_4(a_s,n_f), \\&
    \gamma_{3,3}^{T} = (-3
+2 a_s n_f
) (-4320
+a_s n_f \big(
        3051
        +14 a_s n_f (-51
        +4 a_s n_f
        )
\big))\mathcal{F}_3^T(a_s,n_f), \\&
    \gamma_{3,2}^{T,\D} = -3 (-3
+a_s n_f
) (-3
+2 a_s n_f
) (135
+8 a_s n_f (-6
+a_s n_f
)
)\mathcal{F}_3^T(a_s,n_f),\\&
    \gamma_{3,1}^{T,\D} = 27 (-3
+2 a_s n_f
) (15
+a_s n_f (-7
+2 a_s n_f
))\mathcal{F}_3^T(a_s,n_f),\\&
    \gamma_{3,0}^{T,\D} = -9 (45
+4 a_s n_f (-9
+4 a_s n_f
))\mathcal{F}_3^T(a_s,n_f), \\&
    \gamma_{2,2}^{T} = (3
-2 a_s n_f
) (234
+5 a_s n_f (-27
+4 a_s n_f
))\mathcal{F}_2^T(a_s,n_f), \\&
    \gamma_{2.1}^{T,\D} = 4 (-3
+2 a_s n_f
) (18
+a_s n_f (-9
+2 a_s n_f
)\mathcal{F}_2^T(a_s,n_f), \\&
    \gamma_{2,0}^{T,\D} = -9 (6
+a_s n_f (-5
+2 a_s n_f
)
)\mathcal{F}_2^T(a_s,n_f), \\&
    \gamma_{1,1}^{T} = -3 (-3
+2 a_s n_f
) (-9
+4 a_s n_f
)\mathcal{F}_1^T(a_s,n_f), \\&
    \gamma_{1,0}^{T,\D} = 27
+8 a_s n_f (-3
+a_s n_f)\mathcal{F}_1^T(a_s,n_f), \\&
\label{eq:TLast}
    \gamma_{0,0}^{T} = (9 - 2 a nf) (-3 + 4 a nf))\mathcal{F}_1^T(a_s,n_f).
\end{align}

\subsection{Results in the Gegenbauer basis}
In the Gegenbauer basis, the spin-five ADM is written as
\begin{equation}
    \begin{pmatrix}
        \gamma_{4,4} & 0 & \gamma_{4,2}^{\G} & 0 & \gamma_{4,0}^{\G} \\[6pt]
        0 & \gamma_{3,3} & 0 & \gamma_{3,1}^{\G} & 0 \\[6pt]
        0 & 0 & \gamma_{2,2} & 0 & \gamma_{2,0}^{\G} \\[6pt]
        0 & 0 & 0 & \gamma_{1,1} & 0 \\[6pt]
        0 & 0 & 0 & 0 & \gamma_{0,0}
    \end{pmatrix}.
\end{equation}
The corresponding off-diagonal elements are obtained by applying the basis transformation formula, Eq.(\ref{eq:basisTransIn}), to Eqs.(\ref{eq:wFirst})-(\ref{eq:WLast}) for the Wilson operators and Eqs.(\ref{eq:TFirst})-(\ref{eq:TLast}) for the transversity ones. For the former this leads to
\begin{align}
\label{eq:GFirst}
    &\gamma_{4,2}^{G} =\frac{14}{3} a_s n_f (-3
+a_s n_f
) \big(
        171
        +2 a_s n_f (-33
        +4 a_s n_f
        )
\big)\mathcal{F}_4(a_s,n_f) ,\\&
    \gamma_{4,0}^{\G} = 2 a_s n_f \big(
        -1053
        +144 a_s n_f
        +8 a_s^3 n_f^3
\big)\mathcal{F}_4(a_s,n_f),\\&
    \gamma_{3,1}^{\G} =\frac{10 }{3} a_s n_f\big(
        99
        +2 a_s n_f (-27
        +4 a_s n_f
        )
\big)\mathcal{F}_3(a_s,n_f) ,\\&
    \gamma_{2,0}^{\G} = 2 a_s n_f (-15
+2 a_s n_f
) (-15
+4 a_s n_f
)\mathcal{F}_3(a_s,n_f)
\end{align}
while for the latter we find
\begin{align}
    &\gamma_{4,2}^{T,\G} =28 a_s n_f (-3
+a_s n_f
) (-3
+2 a_s n_f
) (81
+a_s n_f (-27
+4 a_s n_f
))\mathcal{F}_4^T(a_s,n_f) ,\\&
    \gamma_{4,0}^{T,\G} = 6 a_s n_f \big(
        2673
        +a_s n_f \big(
                -2079
                +522 a_s n_f
                +16 a_s^3 n_f^3
        \big)
\big)\mathcal{F}_4^T(a_s,n_f),\\&
    \gamma_{3,1}^{T,\G} =20 a_s n_f (-3
+2 a_s n_f
) (45
+a_s n_f (-21
+4 a_s n_f
))\mathcal{F}_3^T(a_s,n_f) ,\\&
\label{eq:GLast}
    \gamma_{2,0}^{T,\G} = -12 a_s n_f (18
+a_s n_f (-15
+4 a_s n_f
)
)\mathcal{F}_2^T(a_s,n_f).
\end{align}
The results presented in Sec.\ref{sec:largeNF} correspond to the series expansions of Eqs.(\ref{eq:GFirst})-(\ref{eq:GLast}) to sixth order in $a_s$.

\renewcommand{\theequation}{\ref{ap:L2ADM}.\arabic{equation}}
\setcounter{equation}{0}
\renewcommand{\thefigure}{\ref{ap:L2ADM}.\arabic{figure}}
\setcounter{figure}{0}
\renewcommand{\thetable}{\ref{ap:L2ADM}.\arabic{table}}
\setcounter{table}{0}
\section{Two-loop $SU(n_c)$ anomalous dimensions: Wilson operators}
\label{ap:L2ADM}
In this appendix we collect the off-diagonal elements of the two-loop spin-$N$ Wilson ADMs with $N=2,4,6,8,10$ in the derivative basis. The results are presented for an arbitrary color gauge group $SU(n_c)$.
\subsection{Spin two}
\begin{equation}
    \gamma_{1,0}^{\D,(1)} = -\frac{188 }{27}\text{\:\CA} \text{\:\CF}
+\frac{56 }{27}\text{\:\CF}^2
+\frac{32 }{27}\text{\:\CF} \text{\nf}
\end{equation}
This expression agrees with the one presented in \cite{Gracey:2009da}.
\subsection{Spin four}
\begin{align}
    \gamma_{3,2}^{\D,(1)} =& -\frac{1391 }{225}C_A C_F
+\frac{343 }{375}C_F^2
+\frac{362 }{225}C_F n_f \\
    \gamma_{3,1}^{\D,(1)} =& -\frac{211 }{100}C_A C_F
-\frac{1153 }{9000}C_F^2
+\frac{13 }{25}C_F n_f \\
    \gamma_{3,0}^{\D,(1)} =& -\frac{958 }{675}C_A C_F
+\frac{772 }{3375}C_F^2
+\frac{106 }{675}C_F n_f
\end{align}
\subsection{Spin six}
\begin{align}
    \gamma_{5,4}^{\D,(1)} =& -\frac{14477 }{2205}C_A C_F
+\frac{24809 }{25725}C_F^2
+\frac{3896 }{2205}C_F n_f\\
    \gamma_{5,3}^{\D,(1)} =& -\frac{365 }{147}C_A C_F
+\frac{25889 }{77175}C_F^2
+\frac{197 }{294}C_F n_f\\
    \gamma_{5,2}^{\D,(1)} =& -\frac{1640 }{1323}C_A C_F
+\frac{36541 }{926100}C_F^2
+\frac{2084 }{6615}C_F n_f\\
    \gamma_{5,1}^{\D,(1)} =& -\frac{1991 }{3528}C_A C_F
-\frac{51931 }{123480}C_F^2
+\frac{1283 }{8820}C_F n_f\\
    \gamma_{5,0}^{\D,(1)} =& -\frac{503 }{882}C_A C_F
-\frac{58763 }{771750}C_F^2
+\frac{482 }{11025}C_F n_f
\end{align}
\subsection{Spin Eight}
\begin{align}
    \gamma_{7,6}^{\D,(1)} =& -\frac{583733 }{85050}C_A C_F
+\frac{1699543 }{1786050}C_F^2
+\frac{3163 }{1701}C_F n_f\\
    \gamma_{7,5}^{\D,(1)} =& -\frac{267829 }{97200}C_A C_F
+\frac{13682701 }{28576800}C_F^2
+\frac{184 }{243}C_F n_f\\
    \gamma_{7,4}^{\D,(1)} =& -\frac{48971 }{34020}C_A C_F
+\frac{290249 }{1428840}C_F^2
+\frac{3371 }{8505}C_F n_f\\
    \gamma_{7,3}^{\D,(1)} =& -\frac{42251 }{54432}C_A C_F
-\frac{91909 }{1166400}C_F^2
+\frac{1507 }{6804}C_F n_f\\
    \gamma_{7,2}^{\D,(1)} =& -\frac{832 }{1701}C_A C_F
-\frac{4358551 }{35721000}C_F^2
+\frac{5167 }{42525}C_F n_f\\
    \gamma_{7,1}^{\D,(1)} =& -\frac{137293 }{680400}C_A C_F
-\frac{12067103 }{28576800}C_F^2
+\frac{1003 }{17010}C_F n_f\\
    \gamma_{7,0}^{\D,(1)} =&-\frac{347923 }{1190700}C_A C_F
-\frac{1799867 }{12502350}C_F^2
+\frac{851 }{59535}C_F n_f
\end{align}
\subsection{Spin ten}
\begin{align}
    \gamma_{9,8}^{\D,(1)} =& -\frac{2061007 }{291060}C_A C_F
+\frac{6444973 }{7043652}C_F^2
+\frac{31376 }{16335}C_F n_f\\
    \gamma_{9,7}^{\D,(1)} =& -\frac{1913321 }{646800}C_A C_F
+\frac{1127023897 }{2113095600}C_F^2
+\frac{5921 }{7260}C_F n_f\\
    \gamma_{9,6}^{\D,(1)} =& -\frac{251803 }{155925}C_A C_F
+\frac{1878918313 }{6339286800}C_F^2
+\frac{154876 }{343035}C_F n_f\\
    \gamma_{9,5}^{\D,(1)} =& -\frac{224413 }{237600}C_A C_F
+\frac{114422047 }{1408730400}C_F^2
+\frac{35683 }{130680}C_F n_f\\
    \gamma_{9,4}^{\D,(1)} =& -\frac{52399 }{86625}C_A C_F
+\frac{94076813 }{10565478000}C_F^2
+\frac{32251 }{190575}C_F n_f\\
    \gamma_{9,3}^{\D,(1)} =& -\frac{31973}{99792} C_A C_F
-\frac{659890733 }{3169643400}C_F^2
+\frac{141389 }{1372140}C_F n_f\\
    \gamma_{9,2}^{\D,(1)} =& -\frac{178939 }{727650}C_A C_F
-\frac{738295993 }{4930556400}C_F^2
+\frac{46946 }{800415}C_F n_f\\
    \gamma_{9,1}^{\D,(1)} =& -\frac{575083 }{7761600}C_A C_F
-\frac{3198830903 }{8452382400}C_F^2
+\frac{3401 }{121968}C_F n_f\\
    \gamma_{9,0}^{\D,(1)} =& -\frac{4470791 }{26195400}C_A C_F
-\frac{182838473 }{1188616275}C_F^2
+\frac{4069 }{1029105}C_F n_f
\end{align}

\renewcommand{\theequation}{\ref{ap:L2ADMTr}.\arabic{equation}}
\setcounter{equation}{0}
\renewcommand{\thefigure}{\ref{ap:L2ADMTr}.\arabic{figure}}
\setcounter{figure}{0}
\renewcommand{\thetable}{\ref{ap:L2ADMTr}.\arabic{table}}
\setcounter{table}{0}
\section{Two-loop $SU(n_c)$ anomalous dimensions: Transversity operators}
\label{ap:L2ADMTr}
In this appendix we collect the off-diagonal elements of the two-loop spin-$N$ transversity ADMs with $N=2,4,6,8,10$ in the derivative basis. The results are presented for an arbitrary color gauge group $SU(n_c)$.
\subsection{Spin two}
\begin{equation}
    \gamma_{1,0}^{T,\D,(1)} = -\frac{29 }{18}\text{\CA} \text{\CF}
-\frac{5 }{2}C_F^2
+\frac{7 }{9}\text{\CF} \text{\nf}
\end{equation}
This expression agrees with the one presented in \cite{Gracey:2009da}\footnote{The author is grateful to J. Gracey for pointing out an error in this expression in a previous version of this text.}.
\subsection{Spin four}
\begin{align}
    \gamma_{3,2}^{T,\D,(1)} =& -\frac{9 }{2}\text{\CA} \text{\CF}
-\frac{5 }{12}C_F^2
+\frac{17 }{12}\text{\CF} \text{\nf}\\
    \gamma_{3,1}^{T,\D,(1)} =& -\frac{16 }{9}\text{\CA} \text{\CF}
+\frac{13 }{12}C_F^2
+\frac{13 }{36}\text{\CF} \text{\nf}\\
    \gamma_{3,0}^{T,\D,(1)} =& \: \: \frac{47 }{108}\text{\CA} \text{\CF}
-\frac{175 }{108}C_F^2
+\frac{7 }{108}\text{\CF} \text{\nf}
\end{align}
\subsection{Spin six}
\begin{align}
    \gamma_{5,4}^{T,\D,(1)} =& -\frac{113 }{20}\text{\CA} \text{\CF}
+\frac{91 }{540}C_F^2
+\frac{5 }{3}\text{\CF} \text{\nf}\\
    \gamma_{5,3}^{T,\D,(1)} =& -\frac{241 }{90}\text{\CA} \text{\CF}
+\frac{229 }{135}C_F^2
+\frac{26 }{45}\text{\CF} \text{\nf}\\
    \gamma_{5,2}^{T,\D,(1)} =& \: \: \frac{11 }{30}\text{\CA} \text{\CF}
-\frac{12193 }{5400}C_F^2
+\frac{7 }{30}\text{\CF} \text{\nf}\\
    \gamma_{5,1}^{T,\D,(1)} =& -\frac{103 }{108}\text{\CA} \text{\CF}
+\frac{2221 }{1800}C_F^2
+\frac{43 }{540}\text{\CF} \text{\nf}\\
    \gamma_{5,0}^{T,\D,(1)} =& \: \: \frac{2227 }{5400}\text{\CA} \text{\CF}
-\frac{421 }{375}C_F^2
+\frac{13 }{1350}\text{\CF} \text{\nf}
\end{align}
\subsection{Spin Eight}
\begin{align}
    \gamma_{7,6}^{T,\D,(1)} =& -\frac{631651 }{100800}\text{\CA} \text{\CF}
+\frac{38891 }{100800}C_F^2
+\frac{259 }{144}\text{\CF} \text{\nf}\\
    \gamma_{7,5}^{T,\D,(1)} =&-\frac{107893}{33600}\text{\CA} \text{\CF}
+\frac{9539 }{4800}C_F^2
+\frac{235 }{336}\text{\CF} \text{\nf}\\
    \gamma_{7,4}^{T,\D,(1)} =& \: \: \frac{509 }{960}\text{\CA} \text{\CF}
-\frac{268213 }{84672}C_F^2
+\frac{115 }{336}\text{\CF} \text{\nf}\\
    \gamma_{7,3}^{T,\D,(1)} =& -\frac{51049 }{20160}\text{\CA} \text{\CF}
+\frac{2805311 }{705600}C_F^2
+\frac{289 }{1680}\text{\CF} \text{\nf}\\
    \gamma_{7,2}^{T,\D,(1)} =& \: \: \frac{16941 }{11200}\text{\CA} \text{\CF}
-\frac{12715673 }{3528000}C_F^2
+\frac{219 }{2800}\text{\CF} \text{\nf}\\
    \gamma_{7,1}^{T,\D,(1)} =& -\frac{259661 }{302400}\text{\CA} \text{\CF}
+\frac{990103 }{705600}C_F^2
+\frac{377 }{15120}\text{\CF} \text{\nf}\\
    \gamma_{7,0}^{T,\D,(1)} =& \: \: \frac{237019 }{705600}\text{\CA} \text{\CF}
-\frac{1386317 }{1646400}C_F^2
-\frac{43 }{35280}\text{\CF} \text{\nf}
\end{align}
\subsection{Spin ten}
\begin{align}
    \gamma_{9,8}^{T,\D,(1)} =& -\frac{5865893 }{882000}\text{\CA} \text{\CF}
+\frac{420299 }{882000}C_F^2
+\frac{47 }{25}\text{\CF} \text{\nf}\\
    \gamma_{9,7}^{T,\D,(1)} =& -\frac{2353831 }{661500}\text{\CA} \text{\CF}
+\frac{235763 }{110250}C_F^2
+\frac{524 }{675}\text{\CF} \text{\nf}\\
    \gamma_{9,6}^{T,\D,(1)} =& \: \: \frac{971519 }{1134000}\text{\CA} \text{\CF}
-\frac{28973947 }{6804000}C_F^2
+\frac{3353 }{8100}\text{\CF} \text{\nf}\\
    \gamma_{9,5}^{T,\D,(1)} =& -\frac{392261 }{84000}\text{\CA} \text{\CF}
+\frac{31375213 }{3969000}C_F^2
+\frac{2993 }{12600}\text{\CF} \text{\nf}\\
    \gamma_{9,4}^{T,\D,(1)} =& \: \: \frac{413573 }{94500}\text{\CA} \text{\CF}
-\frac{1880517 }{196000}C_F^2
+\frac{2573 }{18900}\text{\CF} \text{\nf}\\
    \gamma_{9,3}^{T,\D,(1)} =& -\frac{588691 }{141750}\text{\CA} \text{\CF}
+\frac{53098909 }{6804000}C_F^2
+\frac{2069 }{28350}\text{\CF} \text{\nf}\\
    \gamma_{9,2}^{T,\D,(1)} =& \: \: \frac{2025061 }{882000}\text{\CA} \text{\CF}
-\frac{544026653 }{111132000}C_F^2
+\frac{1439 }{44100}\text{\CF} \text{\nf}\\
    \gamma_{9,1}^{T,\D,(1)} =& -\frac{9073573 }{10584000}\text{\CA} \text{\CF}
+\frac{16156969 }{10584000}C_F^2
+\frac{599 }{75600}\text{\CF} \text{\nf}\\
    \gamma_{9,0}^{T,\D,(1)} =& \: \: \frac{13125529 }{47628000}\text{\CA} \text{\CF}
-\frac{47609063 }{71442000}C_F^2
-\frac{661 }{170100}\text{\CF} \text{\nf}
\end{align}

\bibliographystyle{JHEP}
\bibliography{omebib}

\providecommand{\href}[2]{#2}\begingroup\raggedright\begin{thebibliography}{10}

\bibitem{Muller:1994ses}
D.~M\"uller, D.~Robaschik, B.~Geyer, F.M.~Dittes and J.~Ho\v{r}ej\v{s}i,
  \emph{{Wave functions, evolution equations and evolution kernels from light
  ray operators of QCD}},
  \href{https://dx.doi.org/10.1002/prop.2190420202}{\emph{Fortsch. Phys.} {\bf
  42} (1994) 101} [\href{https://arxiv.org/abs/hep-ph/9812448}{{\tt
  hep-ph/9812448}}].

\bibitem{Ji:1996ek}
X.D.~Ji, \emph{{Gauge-Invariant Decomposition of Nucleon Spin}},
  \href{https://dx.doi.org/10.1103/PhysRevLett.78.610}{\emph{Phys. Rev. Lett.}
  {\bf 78} (1997) 610} [\href{https://arxiv.org/abs/hep-ph/9603249}{{\tt
  hep-ph/9603249}}].

\bibitem{Ji:1996nm}
X.D.~Ji, \emph{{Deeply virtual Compton scattering}},
  \href{https://dx.doi.org/10.1103/PhysRevD.55.7114}{\emph{Phys. Rev. D} {\bf
  55} (1997) 7114} [\href{https://arxiv.org/abs/hep-ph/9609381}{{\tt
  hep-ph/9609381}}].

\bibitem{Radyushkin:1996nd}
A.V.~Radyushkin, \emph{{Scaling limit of deeply virtual Compton scattering}},
  \href{https://dx.doi.org/10.1016/0370-2693(96)00528-X}{\emph{Phys. Lett. B}
  {\bf 380} (1996) 417} [\href{https://arxiv.org/abs/hep-ph/9604317}{{\tt
  hep-ph/9604317}}].

\bibitem{Radyushkin:1996ru}
A.V.~Radyushkin, \emph{{Asymmetric gluon distributions and hard diffractive
  electroproduction}},
  \href{https://dx.doi.org/10.1016/0370-2693(96)00844-1}{\emph{Phys. Lett. B}
  {\bf 385} (1996) 333} [\href{https://arxiv.org/abs/hep-ph/9605431}{{\tt
  hep-ph/9605431}}].

\bibitem{Diehl:2003ny}
M.~Diehl, \emph{{Generalized parton distributions}},
  \href{https://dx.doi.org/10.1016/j.physrep.2003.08.002}{\emph{Phys. Rept.}
  {\bf 388} (2003) 41} [\href{https://arxiv.org/abs/hep-ph/0307382}{{\tt
  hep-ph/0307382}}].

\bibitem{Aidala:2012mv}
C.A.~Aidala, S.D.~Bass, D.~Hasch and G.K.~Mallot, \emph{{The Spin Structure of
  the Nucleon}}, \href{https://dx.doi.org/10.1103/RevModPhys.85.655}{\emph{Rev.
  Mod. Phys.} {\bf 85} (2013) 655} [\href{https://arxiv.org/abs/1209.2803}{{\tt
  arXiv:1209.2803}}].

\bibitem{Leader:2013jra}
E.~Leader and C.~Lorc\'e, \emph{{The angular momentum controversy:
  What\textquoteright{}s it all about and does it matter?}},
  \href{https://dx.doi.org/10.1016/j.physrep.2014.02.010}{\emph{Phys. Rept.}
  {\bf 541} (2014) 163} [\href{https://arxiv.org/abs/1309.4235}{{\tt
  arXiv:1309.4235}}].

\bibitem{Deur:2018roz}
A.~Deur, S.J.~Brodsky and G.F.~De~T\'eramond, \emph{{The Spin Structure of the
  Nucleon}},  \href{https://arxiv.org/abs/1807.05250}{{\tt arXiv:1807.05250}}.

\bibitem{Ji:2020ena}
X.~Ji, F.~Yuan and Y.~Zhao, \emph{{What we know and what we
  don\textquoteright{}t know about the proton spin after 30 years}},
  \href{https://dx.doi.org/10.1038/s42254-020-00248-4}{\emph{Nature Rev. Phys.}
  {\bf 3} (2021) 27} [\href{https://arxiv.org/abs/2009.01291}{{\tt
  arXiv:2009.01291}}].

\bibitem{Abramowicz:2015mha}
{\scshape H1, ZEUS} collaboration, H.~Abramowicz et~al., \emph{{Combination of
  measurements of inclusive deep inelastic ${e^{\pm }p}$ scattering cross
  sections and QCD analysis of HERA data}},
  \href{https://dx.doi.org/10.1140/epjc/s10052-015-3710-4}{\emph{Eur. Phys. J.
  C} {\bf 75} (2015) 580} [\href{https://arxiv.org/abs/1506.06042}{{\tt
  arXiv:1506.06042}}].

\bibitem{Accardi:2016ndt}
A.~Accardi et~al., \emph{{A Critical Appraisal and Evaluation of Modern PDFs}},
  \href{https://dx.doi.org/10.1140/epjc/s10052-016-4285-4}{\emph{Eur. Phys. J.
  C} {\bf 76} (2016) 471} [\href{https://arxiv.org/abs/1603.08906}{{\tt
  arXiv:1603.08906}}].

\bibitem{Boer:2011fh}
D.~Boer et~al., \emph{{Gluons and the quark sea at high energies:
  Distributions, polarization, tomography}},
  \href{https://arxiv.org/abs/1108.1713}{{\tt arXiv:1108.1713}}.

\bibitem{AbdulKhalek:2021gbh}
R.~Abdul~Khalek et~al., \emph{{Science Requirements and Detector Concepts for
  the Electron-Ion Collider: EIC Yellow Report}},
  \href{https://arxiv.org/abs/2103.05419}{{\tt arXiv:2103.05419}}.

\bibitem{Gross:1973ju}
D.J.~Gross and F.~Wilczek, \emph{{Asymptotically Free Gauge Theories - I}},
  \href{https://dx.doi.org/10.1103/PhysRevD.8.3633}{\emph{Phys. Rev. D} {\bf 8}
  (1973) 3633}.

\bibitem{Floratos:1977au}
E.G.~Floratos, D.A.~Ross and C.T.~Sachrajda, \emph{{Higher Order Effects in
  Asymptotically Free Gauge Theories: The Anomalous Dimensions of Wilson
  Operators}},
  \href{https://dx.doi.org/10.1016/0550-3213(77)90020-7}{\emph{Nucl. Phys. B}
  {\bf 129} (1977) 66}.

\bibitem{Moch:2004pa}
S.~Moch, J.A.M.~Vermaseren and A.~Vogt, \emph{{The Three loop splitting
  functions in QCD: The Nonsinglet case}},
  \href{https://dx.doi.org/10.1016/j.nuclphysb.2004.03.030}{\emph{Nucl. Phys.
  B} {\bf 688} (2004) 101} [\href{https://arxiv.org/abs/hep-ph/0403192}{{\tt
  hep-ph/0403192}}].

\bibitem{Blumlein:2021enk}
J.~Bl{\"u}mlein, P.~Marquard, C.~Schneider and K.~Sch\"onwald, \emph{{The
  three-loop unpolarized and polarized non-singlet anomalous dimensions from
  off shell operator matrix elements}},
  \href{https://dx.doi.org/10.1016/j.nuclphysb.2021.115542}{\emph{Nucl. Phys.
  B} {\bf 971} (2021) 115542} [\href{https://arxiv.org/abs/2107.06267}{{\tt
  arXiv:2107.06267}}].

\bibitem{Gracey:1994nn}
J.A.~Gracey, \emph{{Anomalous dimension of nonsinglet Wilson operators at
  $O(1/n_f)$ in deep inelastic scattering}},
  \href{https://dx.doi.org/10.1016/0370-2693(94)90502-9}{\emph{Phys. Lett. B}
  {\bf 322} (1994) 141} [\href{https://arxiv.org/abs/hep-ph/9401214}{{\tt
  hep-ph/9401214}}].

\bibitem{Davies:2016jie}
J.~Davies, A.~Vogt, B.~Ruijl, T.~Ueda and J.A.M.~Vermaseren, \emph{{Large-$n_f$
  contributions to the four-loop splitting functions in QCD}},
  \href{https://dx.doi.org/10.1016/j.nuclphysb.2016.12.012}{\emph{Nucl. Phys.
  B} {\bf 915} (2017) 335} [\href{https://arxiv.org/abs/1610.07477}{{\tt
  arXiv:1610.07477}}].

\bibitem{Velizhanin:2011es}
V.N.~Velizhanin, \emph{{Four loop anomalous dimension of the second moment of
  the non-singlet twist-2 operator in QCD}},
  \href{https://dx.doi.org/10.1016/j.nuclphysb.2012.03.006}{\emph{Nucl. Phys.
  B} {\bf 860} (2012) 288} [\href{https://arxiv.org/abs/1112.3954}{{\tt
  arXiv:1112.3954}}].

\bibitem{Velizhanin:2014fua}
V.N.~Velizhanin, \emph{{Four-loop anomalous dimension of the third and fourth
  moments of the nonsinglet twist-2 operator in QCD}},
  \href{https://dx.doi.org/10.1142/S0217751X20501997}{\emph{Int. J. Mod. Phys.
  A} {\bf 35} (2020) 2050199} [\href{https://arxiv.org/abs/1411.1331}{{\tt
  arXiv:1411.1331}}].

\bibitem{Ruijl:2016pkm}
B.~Ruijl, T.~Ueda, J.A.M.~Vermaseren, J.~Davies and A.~Vogt, \emph{{First
  Forcer results on deep-inelastic scattering and related quantities}},
  \href{https://dx.doi.org/10.22323/1.260.0071}{\emph{PoS} {\bf LL2016} (2016)
  071} [\href{https://arxiv.org/abs/1605.08408}{{\tt arXiv:1605.08408}}].

\bibitem{Moch:2017uml}
S.~Moch, B.~Ruijl, T.~Ueda, J.A.M.~Vermaseren and A.~Vogt, \emph{{Four-Loop
  Non-Singlet Splitting Functions in the Planar Limit and Beyond}},
  \href{https://dx.doi.org/10.1007/JHEP10(2017)041}{\emph{JHEP} {\bf 10} (2017)
  041} [\href{https://arxiv.org/abs/1707.08315}{{\tt arXiv:1707.08315}}].

\bibitem{Herzog:2018kwj}
F.~Herzog, S.~Moch, B.~Ruijl, T.~Ueda, J.A.M.~Vermaseren and A.~Vogt,
  \emph{{Five-loop contributions to low-N non-singlet anomalous dimensions in
  QCD}}, \href{https://dx.doi.org/10.1016/j.physletb.2019.01.060}{\emph{Phys.
  Lett. B} {\bf 790} (2019) 436} [\href{https://arxiv.org/abs/1812.11818}{{\tt
  arXiv:1812.11818}}].

\bibitem{Blumlein:2023aso}
J.~Bl\"umlein, \emph{{Deep-Inelastic Scattering: What do we know ?}},
  \href{https://arxiv.org/abs/2306.01362}{{\tt arXiv:2306.01362}}.

\bibitem{Braun:2003rp}
V.M.~Braun, G.P.~Korchemsky and D.~M\"uller, \emph{{The Uses of conformal
  symmetry in QCD}},
  \href{https://dx.doi.org/10.1016/S0146-6410(03)90004-4}{\emph{Prog. Part.
  Nucl. Phys.} {\bf 51} (2003) 311}
  [\href{https://arxiv.org/abs/hep-ph/0306057}{{\tt hep-ph/0306057}}].

\bibitem{Braun:2017cih}
V.M.~Braun, A.N.~Manashov, S.~Moch and M.~Strohmaier, \emph{{Three-loop
  evolution equation for flavor-nonsinglet operators in off-forward
  kinematics}}, \href{https://dx.doi.org/10.1007/JHEP06(2017)037}{\emph{JHEP}
  {\bf 06} (2017) 037} [\href{https://arxiv.org/abs/1703.09532}{{\tt
  arXiv:1703.09532}}].

\bibitem{Mueller:1991gd}
D.~M{\"u}ller, \emph{{Constraints for anomalous dimensions of local light cone
  operators in $\phi^3$ in six-dimensions theory}},
  \href{https://dx.doi.org/10.1007/BF01555504}{\emph{Z. Phys. C} {\bf 49}
  (1991) 293}.

\bibitem{Braun:2016qlg}
V.M.~Braun, A.N.~Manashov, S.~Moch and M.~Strohmaier, \emph{{Two-loop conformal
  generators for leading-twist operators in QCD}},
  \href{https://dx.doi.org/10.1007/JHEP03(2016)142}{\emph{JHEP} {\bf 03} (2016)
  142} [\href{https://arxiv.org/abs/1601.05937}{{\tt arXiv:1601.05937}}].

\bibitem{Artru:1989zv}
X.~Artru and M.~Mekhfi, \emph{{Transversely Polarized Parton Densities, their
  Evolution and their Measurement}},
  \href{https://dx.doi.org/10.1007/BF01556280}{\emph{Z. Phys. C} {\bf 45}
  (1990) 669}.

\bibitem{Shifman:1980dk}
M.A.~Shifman and M.I.~Vysotsky, \emph{{Form-Factors of Heavy Mesons in QCD}},
  \href{https://dx.doi.org/10.1016/0550-3213(81)90023-7}{\emph{Nucl. Phys. B}
  {\bf 186} (1981) 475}.

\bibitem{Baldracchini:1981}
F.~Baldracchini, N.S.~Craigie, V.~Roberto and M.~Socolovsky, \emph{{A Survey of
  Polarization Asymmetries Predicted by QCD}},
  \href{https://dx.doi.org/https://doi.org/10.1002/prop.19810291102}{\emph{Fortschritte
  der Physik} {\bf 29} (1981) 505}.

\bibitem{Geyer:1982fk}
B.~Geyer, \emph{{Anomalous dimensions in local and non-local light cone
  expansion. (Talk)}},
  \href{https://dx.doi.org/10.1007/BF01596709}{\emph{Czech. J. Phys. B} {\bf
  32} (1982) 645}.

\bibitem{Blumlein:1999sc}
J.~Bl{\"u}mlein, B.~Geyer and D.~Robaschik, \emph{{The Virtual Compton
  amplitude in the generalized Bjorken region: twist-2 contributions}},
  \href{https://dx.doi.org/10.1016/S0550-3213(99)00418-6}{\emph{Nucl. Phys. B}
  {\bf 560} (1999) 283} [\href{https://arxiv.org/abs/hep-ph/9903520}{{\tt
  hep-ph/9903520}}].

\bibitem{Blumlein:2001ca}
J.~Bl{\"u}mlein, \emph{{On the anomalous dimension of the transversity
  distribution h$_1$(x,$Q^2$)}},
  \href{https://dx.doi.org/10.1007/s100520100703}{\emph{Eur. Phys. J. C} {\bf
  20} (2001) 683} [\href{https://arxiv.org/abs/hep-ph/0104099}{{\tt
  hep-ph/0104099}}].

\bibitem{Gracey:2009da}
J.A.~Gracey, \emph{{Three loop anti-MS operator correlation functions for deep
  inelastic scattering in the chiral limit}},
  \href{https://dx.doi.org/10.1088/1126-6708/2009/04/127}{\emph{JHEP} {\bf 04}
  (2009) 127} [\href{https://arxiv.org/abs/0903.4623}{{\tt arXiv:0903.4623}}].

\bibitem{Kniehl:2020nhw}
B.A.~Kniehl and O.L.~Veretin, \emph{{Moments $n=2$ and $n=3$ of the Wilson
  twist-two operators at three loops in the RI${}'$/SMOM scheme}},
  \href{https://dx.doi.org/10.1016/j.nuclphysb.2020.115229}{\emph{Nucl. Phys.
  B} {\bf 961} (2020) 115229} [\href{https://arxiv.org/abs/2009.11325}{{\tt
  arXiv:2009.11325}}].

\bibitem{Moch:2021cdq}
S.~Moch and S.~Van~Thurenhout, \emph{{Renormalization of non-singlet quark
  operator matrix elements for off-forward hard scattering}},
  \href{https://dx.doi.org/10.1016/j.nuclphysb.2021.115536}{\emph{Nucl. Phys.
  B} {\bf 971} (2021) 115536} [\href{https://arxiv.org/abs/2107.02470}{{\tt
  arXiv:2107.02470}}].

\bibitem{VanThurenhout:2022nmx}
S.~Van~Thurenhout, \emph{{Off-forward anomalous dimensions of non-singlet
  transversity operators}},
  \href{https://dx.doi.org/10.1016/j.nuclphysb.2022.115835}{\emph{Nucl. Phys.
  B} {\bf 980} (2022) 115835} [\href{https://arxiv.org/abs/2204.02140}{{\tt
  arXiv:2204.02140}}].

\bibitem{VanThurenhout:2022hgd}
S.~Van~Thurenhout and S.O.~Moch, \emph{{Off-forward anomalous dimensions in the
  leading-$n_f$ limit}},
  \href{https://dx.doi.org/10.22323/1.416.0076}{\emph{PoS} {\bf LL2022} (2022)
  076} [\href{https://arxiv.org/abs/2206.04517}{{\tt arXiv:2206.04517}}].

\bibitem{Gracey:2003mr}
J.A.~Gracey, \emph{{Three loop anomalous dimension of the second moment of the
  transversity operator in the MS-bar and RI-prime schemes}},
  \href{https://dx.doi.org/10.1016/S0550-3213(03)00543-1}{\emph{Nucl. Phys. B}
  {\bf 667} (2003) 242} [\href{https://arxiv.org/abs/hep-ph/0306163}{{\tt
  hep-ph/0306163}}].

\bibitem{Anselmino:1994gn}
M.~Anselmino, A.~Efremov and E.~Leader, \emph{{The Theory and phenomenology of
  polarized deep inelastic scattering}},
  \href{https://dx.doi.org/10.1016/0370-1573(95)00011-5}{\emph{Phys. Rept.}
  {\bf 261} (1995) 1} [\href{https://arxiv.org/abs/hep-ph/9501369}{{\tt
  hep-ph/9501369}}].

\bibitem{Liang:2000gz}
Z.t.~Liang and C.~Boros, \emph{{Single spin asymmetries in inclusive
  high-energy hadron hadron collision processes}},
  \href{https://dx.doi.org/10.1142/S0217751X0000046X}{\emph{Int. J. Mod. Phys.
  A} {\bf 15} (2000) 927} [\href{https://arxiv.org/abs/hep-ph/0001330}{{\tt
  hep-ph/0001330}}].

\bibitem{Barone:2001sp}
V.~Barone, A.~Drago and P.G.~Ratcliffe, \emph{{Transverse polarisation of
  quarks in hadrons}},
  \href{https://dx.doi.org/10.1016/S0370-1573(01)00051-5}{\emph{Phys. Rept.}
  {\bf 359} (2002) 1} [\href{https://arxiv.org/abs/hep-ph/0104283}{{\tt
  hep-ph/0104283}}].

\bibitem{Braun:2009mi}
V.M.~Braun, A.N.~Manashov and B.~Pirnay, \emph{{Scale dependence of twist-three
  contributions to single spin asymmetries}},
  \href{https://dx.doi.org/10.1103/PhysRevD.80.114002}{\emph{Phys. Rev. D} {\bf
  80} (2009) 114002} [\href{https://arxiv.org/abs/0909.3410}{{\tt
  arXiv:0909.3410}}].

\bibitem{Braun:2022gzl}
V.M.~Braun, \emph{{Higher Twists}},
  \href{https://dx.doi.org/10.1051/epjconf/202227401012}{\emph{EPJ Web Conf.}
  {\bf 274} (2022) 01012} [\href{https://arxiv.org/abs/2212.02887}{{\tt
  arXiv:2212.02887}}].

\bibitem{Efremov:1978rn}
A.V.~Efremov and A.V.~Radyushkin, \emph{{Asymptotical Behavior of Pion
  Electromagnetic Form-Factor in QCD}},
  \href{https://dx.doi.org/10.1007/BF01032111}{\emph{Theor. Math. Phys.} {\bf
  42} (1980) 97}.

\bibitem{Efremov:1979qk}
A.V.~Efremov and A.V.~Radyushkin, \emph{{Factorization and Asymptotical
  Behavior of Pion Form-Factor in QCD}},
  \href{https://dx.doi.org/10.1016/0370-2693(80)90869-2}{\emph{Phys. Lett. B}
  {\bf 94} (1980) 245}.

\bibitem{Lepage:1979zb}
G.P.~Lepage and S.J.~Brodsky, \emph{{Exclusive Processes in Quantum
  Chromodynamics: Evolution Equations for Hadronic Wave Functions and the
  Form-Factors of Mesons}},
  \href{https://dx.doi.org/10.1016/0370-2693(79)90554-9}{\emph{Phys. Lett. B}
  {\bf 87} (1979) 359}.

\bibitem{Lepage:1980fj}
G.P.~Lepage and S.J.~Brodsky, \emph{{Exclusive Processes in Perturbative
  Quantum Chromodynamics}},
  \href{https://dx.doi.org/10.1103/PhysRevD.22.2157}{\emph{Phys. Rev. D} {\bf
  22} (1980) 2157}.

\bibitem{Dittes:1988xz}
F.M.~Dittes, D.~M{\"u}ller, D.~Robaschik, B.~Geyer and J.~Ho\v{r}ej\v{s}i,
  \emph{{The Altarelli-Parisi Kernel as Asymptotic Limit of an Extended
  Brodsky-Lepage Kernel}},
  \href{https://dx.doi.org/10.1016/0370-2693(88)90955-0}{\emph{Phys. Lett. B}
  {\bf 209} (1988) 325}.

\bibitem{Belitsky:1998gc}
A.V.~Belitsky and D.~M{\"u}ller, \emph{{Broken conformal invariance and
  spectrum of anomalous dimensions in QCD}},
  \href{https://dx.doi.org/10.1016/S0550-3213(98)00677-4}{\emph{Nucl. Phys. B}
  {\bf 537} (1999) 397} [\href{https://arxiv.org/abs/hep-ph/9804379}{{\tt
  hep-ph/9804379}}].

\bibitem{Gockeler:2004wp}
{\scshape QCDSF} collaboration, M.~G{\"o}ckeler, R.~Horsley, D.~Pleiter,
  P.E.L.~Rakow and G.~Schierholz, \emph{{A Lattice determination of moments of
  unpolarised nucleon structure functions using improved Wilson fermions}},
  \href{https://dx.doi.org/10.1103/PhysRevD.71.114511}{\emph{Phys. Rev. D} {\bf
  71} (2005) 114511} [\href{https://arxiv.org/abs/hep-ph/0410187}{{\tt
  hep-ph/0410187}}].

\bibitem{olver10}
F.W.J.~Olver, D.W.~Lozier, R.F.~Boisvert and C.W.~Clark, \emph{The {NIST}
  Handbook of Mathematical Functions}, Cambridge Univ. Press (2010).

\bibitem{Gracey:2011zn}
J.A.~Gracey, \emph{{Two loop renormalization of the n = 2 Wilson operator in
  the RI'/SMOM scheme}},
  \href{https://dx.doi.org/10.1007/JHEP03(2011)109}{\emph{JHEP} {\bf 03} (2011)
  109} [\href{https://arxiv.org/abs/1103.2055}{{\tt arXiv:1103.2055}}].

\bibitem{Gracey:2011zg}
J.A.~Gracey, \emph{{Amplitudes for the n = 3 moment of the Wilson operator at
  two loops in the RI/'SMOM scheme}},
  \href{https://dx.doi.org/10.1103/PhysRevD.84.016002}{\emph{Phys. Rev. D} {\bf
  84} (2011) 016002} [\href{https://arxiv.org/abs/1105.2138}{{\tt
  arXiv:1105.2138}}].

\bibitem{gradshteyn2007}
I.S.~Gradshteyn and I.M.~Ryzhik, \emph{Table of integrals, series, and
  products}, seventh~ed., Elsevier/Academic Press, Amsterdam (2007).

\bibitem{Braun:2022byg}
V.M.~Braun, K.G.~Chetyrkin and A.N.~Manashov, \emph{{NNLO anomalous dimension
  matrix for twist-two flavor-singlet operators}},
  \href{https://dx.doi.org/10.1016/j.physletb.2022.137409}{\emph{Phys. Lett. B}
  {\bf 834} (2022) 137409} [\href{https://arxiv.org/abs/2205.08228}{{\tt
  arXiv:2205.08228}}].

\bibitem{Makeenko:1980bh}
Y.M.~Makeenko, \emph{{Conformal Operators in Quantum Chromodynamics}},
  {\emph{Sov. J. Nucl. Phys.} {\bf 33} (1981) 440}.

\bibitem{Vermaseren:1998uu}
J.A.M.~Vermaseren, \emph{{Harmonic sums, Mellin transforms and integrals}},
  \href{https://dx.doi.org/10.1142/S0217751X99001032}{\emph{Int. J. Mod. Phys.
  A} {\bf 14} (1999) 2037} [\href{https://arxiv.org/abs/hep-ph/9806280}{{\tt
  hep-ph/9806280}}].

\bibitem{Blumlein:1998if}
J.~Bl{\"u}mlein and S.~Kurth, \emph{{Harmonic sums and Mellin transforms up to
  two loop order}},
  \href{https://dx.doi.org/10.1103/PhysRevD.60.014018}{\emph{Phys. Rev. D} {\bf
  60} (1999) 014018} [\href{https://arxiv.org/abs/hep-ph/9810241}{{\tt
  hep-ph/9810241}}].

\bibitem{Schneider2007}
C.~Schneider, \emph{Symbolic summation assists combinatorics}, {\emph{Seminaire
  Lotharingien de Combinatoire} {\bf 56} (01, 2007) 1}.

\bibitem{Schneider:2013uan}
C.~Schneider, \emph{{Simplifying Multiple Sums in Difference Fields}},  in
  \emph{{LHCPhenoNet School}: {Integration, Summation and Special Functions in
  Quantum Field Theory}}, p.~325, 2013.
\newblock \href{https://dx.doi.org/10.1007/978-3-7091-1616-6_14}{DOI}.

\bibitem{Mikhailov:1984ii}
S.V.~Mikhailov and A.V.~Radyushkin, \emph{{Evolution Kernels in {QCD}: Two Loop
  Calculation in Feynman Gauge}},
  \href{https://dx.doi.org/10.1016/0550-3213(85)90213-5}{\emph{Nucl. Phys. B}
  {\bf 254} (1985) 89}.

\bibitem{Braun:2014vba}
V.M.~Braun and A.N.~Manashov, \emph{{Two-loop evolution equations for light-ray
  operators}},
  \href{https://dx.doi.org/10.1016/j.physletb.2014.05.037}{\emph{Phys. Lett. B}
  {\bf 734} (2014) 137} [\href{https://arxiv.org/abs/1404.0863}{{\tt
  arXiv:1404.0863}}].

\bibitem{Vermaseren:2000nd}
J.A.M.~Vermaseren, \emph{{New features of FORM}},
  \href{https://arxiv.org/abs/math-ph/0010025}{{\tt math-ph/0010025}}.

\bibitem{Kuipers:2012rf}
J.~Kuipers, T.~Ueda, J.A.M.~Vermaseren and J.~Vollinga, \emph{{FORM version
  4.0}}, \href{https://dx.doi.org/10.1016/j.cpc.2012.12.028}{\emph{Comput.
  Phys. Commun.} {\bf 184} (2013) 1453}
  [\href{https://arxiv.org/abs/1203.6543}{{\tt arXiv:1203.6543}}].

\bibitem{Balitsky:1987bk}
I.I.~Balitsky and V.M.~Braun, \emph{{Evolution Equations for QCD String
  Operators}},
  \href{https://dx.doi.org/10.1016/0550-3213(89)90168-5}{\emph{Nucl. Phys. B}
  {\bf 311} (1989) 541}.

\bibitem{Braunschweig:1987dr}
T.~Braunschweig, B.~Geyer and D.~Robaschik, \emph{{Anomalous Dimensions of
  Flavor Singlet Light Cone Operators}}, {\emph{Annalen Phys.} {\bf 44} (1987)
  403}.

\end{thebibliography}\endgroup

\end{document}